\documentclass[aps,nofootinbib,preprintnumbers]{revtex4}
\usepackage{lipsum}
\usepackage{CJK}
\usepackage{amsfonts}
\usepackage{amsmath}
\usepackage{amssymb}
\usepackage[english]{babel}
\usepackage{graphicx}
\usepackage{epsfig}
\usepackage{bm}
\usepackage{verbatim}
\usepackage[utf8]{inputenc}
\usepackage{booktabs}
\usepackage{multirow}
\usepackage{subfig}
\usepackage{slashed}
\usepackage{xcolor}
\usepackage[colorlinks=true,urlcolor=red,citecolor=red]{hyperref}
\usepackage[font=small]{caption}
\usepackage{float}
\usepackage{blindtext}
\usepackage{placeins}
\usepackage[utf8]{inputenc}
\usepackage{bbm}
\usepackage[colorinlistoftodos]{todonotes}

\newenvironment{Eqnarray}{\arraycolsep 0.14em\begin{eqnarray}}{\end{eqnarray}}
\newcommand{\ba}{\begin{Eqnarray}}
\newcommand{\ea}{\end{Eqnarray}}
\newcommand{\be}{\begin{equation}}
\newcommand{\ee}{\end{equation}}
\newcommand{\bal}{\begin{aligned}}
\newcommand{\eal}{\end{aligned}}
\newcommand{\bea}{\begin{eqnarray}}
\newcommand{\eea}{\end{eqnarray}}
\newcommand{\ben}{\begin{enumerate}}
\newcommand{\een}{\end{enumerate}}
\newcommand{\bit}{\begin{itemize}}
\newcommand{\eit}{\end{itemize}}
\newcommand{\bde}{\begin{widetext}}
\newcommand{\ede}{\end{widetext}}

\def\lsim{\mathrel{\rlap{\lower4pt\hbox{\hskip1pt$\sim$}}
    \raise1pt\hbox{$<$}}}
\def\gsim{\mathrel{\rlap{\lower4pt\hbox{\hskip1pt$\sim$}}
    \raise1pt\hbox{$>$}}}

\def\3211{$\mathrm{SU(3) \otimes SU(2)_L \otimes U(1)_R \otimes U(1)_{B-L}}$ }
\def\321{$\mathrm{SU(3) \otimes SU(2) \otimes U(1)}$ }
\def\422{$\mathrm{SU(4) \otimes SU(2) \otimes SU(2)_R}$ }

\newcommand{\mathsym}[1]{{}}

\definecolor{bostonuniversityred}{rgb}{0.8, 0.0, 0.0}

\input{tcilatex}
\begin{document}

\title{Models of radiative linear seesaw with electrically charged mediators}
\author{A. E. C\'arcamo Hern\'andez$^{a,b,c}$}
\email{antonio.carcamo@usm.cl}
\author{Yocelyne Hidalgo Vel\'{a}squez$^a$}
\email{yocehidalgov@gmail.com}
\author{Sergey Kovalenko$^{c,d}$}
\email{sergey.kovalenko@unab.cl}
\author{Nicol\'{a}s A. P\'{e}rez-Julve$^a$}
\email{nicolasperezjulve@gmail.com}
\author{Ivan Schmidt$^{a,b}$}
\email{Deceased}
\affiliation{\hspace{2cm}\\
$^{{a}}$Universidad T\'ecnica Federico Santa Mar\'{\i}a, Casilla 110-V,
Valpara\'{\i}so, Chile\\
$^{{b}}$Centro Cient\'{\i}fico-Tecnol\'ogico de Valpara\'{\i}so, Casilla
110-V, Valpara\'{\i}so, Chile\\
$^{{c}}$Millennium Institute for Subatomic Physics at the High-Energy
Frontier, SAPHIR, Chile\\
$^{{d}}$Universidad Andr\'es s Bello, Facultad de Ciencias
Exactas, \\
Departamento de Ciencias F\'{\i}sicas-Center for Theoretical and
Experimental Particle Physics, \\
Fern\'andez Concha 700, Santiago, Chile }
\date{\today }

\begin{abstract}
We propose two versions of radiative linear seesaw models, where
electrically charged scalars and vector-like leptons generate the Dirac
neutrino mass submatrix at one and two loop levels. In these models, the SM
charged lepton masses are generated from a one loop level radiative seesaw
mechanism mediated by charged exotic vector-like leptons and electrically
neutral scalars running in 
the loops. These models can successfully accommodate the current amount of
dark matter and baryon asymmetries observed in the Universe, as well as the
muon anomalous magnetic moment.\newline
\textbf{Our friend and collaborator Iv\'an Schmidt passed away during the
completion of this work. He will be sorely missed.}
\end{abstract}

\maketitle


\section{Introduction}

\label{intro} The origin of the SM charged fermion mass hierarchy, the tiny
active neutrino masses and the current amount of dark matter relic density
and lepton asymmetries observed in the Universe are one of the most relevant
open issues not addressed by the Standard Model (SM) of Particle Physics.
Several theories have been proposed in order to explain the tiny values of
the light active neutrino masses; see \textit{e.g.} Ref.~\cite{Cai:2017jrq}
for a review and \cite{Arbelaez:2022ejo} for a comprehensive study of one
loop radiative neutrino mass models. The most economical way to generate the
tiny masses of the light active neutrinos, considering the SM gauge
symmetry, is by adding two right-handed Majorana neutrinos that mix with the
light active neutrinos, thus triggering a canonical seesaw mechanism \cite%
{Minkowski:1977sc,Yanagida:1979as,Glashow:1979nm,Mohapatra:1979ia,Gell-Mann:1979vob, Schechter:1980gr, Schechter:1981cv}%
, where either the right handed Majorana neutrinos have to be extremely
heavy, with masses of the order of the Grand Unification scale, or they can
be around the TeV scale thus implying that the Dirac Yukawa couplings have
to be very tiny. In both scenarios, the mixing between the active and
sterile neutrinos is very tiny, leading to strongly suppressed charged
lepton flavor (CLFV) violating signatures, several orders of magnitude below
the experimental sensitivity, thus making this scenario untestable via CLFV
decays. 
One interesting and testable explanation for the tiny masses of
light active neutrinos is the so-called linear seesaw mechanism \cite%
{Mohapatra:1986bd,Akhmedov:1995ip,Akhmedov:1995vm,Malinsky:2005bi,Hirsch:2009mx,Dib:2014fua,Chakraborty:2014hfa,Sinha:2015ooa,Wang:2015saa,CarcamoHernandez:2017cwi,CarcamoHernandez:2019iwh,CarcamoHernandez:2021tlv,Batra:2023ssq,Batra:2023mds,Batra:2023bqj,CarcamoHernandez:2023atk}, 
in which  the masses of light active neutrinos feature a linear dependence
on the Dirac neutrino mass submatrix. 
In the linear seesaw realizations
the mixing between active and sterile neutrinos is several orders of
magnitudes larger than in the type I seesaw, thus resulting in
charged lepton flavor violating decay rates within the reach of experimental sensitivity. Furthermore, the linear seesaw realizations, due to the small
mass splitting between the heavy pseudoDirac neutral leptons, provides a
successfull scenario for resonant leptogenesis.

In the present paper we propose a radiative realization of the linear seesaw
mechanism, where the Dirac neutrino mass submatrix is generated at one or
two loop level. The layout of the remainder of the paper is as follows. In
section \ref{model} we describe two radiative linear seesaw models, where
the Dirac neutrino mass matrix is generated at one and two loops. The
implications of the two loop radiative linear seesaw model in leptogenesis is discussed in section \ref{lepto}. The consequences of the radiative linear seesaw models in muon anomalous magnetic moment, dark matter, charged lepton flavor violation and leptogenesis are discusssed in sections \ref{sec:G-2}, \ref{section-DM}, \ref{clfv} and \ref{lepto}, respectively. We state our conclusions in section \ref%
{conclusions}. Appendix \ref{fullnumatrix} shows in full detail the
perturbative diagonalization procedure of the full $7\times 7$ neutrino mass
matrix.

\section{The models}

\label{model} In this section we discuss two radiative linear seesaw models
where the Dirac neutrino mass matrix is generated at one and two loops from
the virtual exchange of electrically charged mediators. We discuss the
phenomenological consequences of these models for the muon anomalous
magnetic moment. Before describing two radiative linear seesaw models, where
the Dirac neutrino mass matrix is generated at one and two loops, we start
by explaining the motivations behind the inclusion of extra scalars,
fermions and symmetries needed for implementing the linear seesaw mechanism
at one and two loop levels and to generate the SM charged lepton masses at
one loop level. The masses of the active light neutrinos arise from a linear
seesaw mechanism when the full neutrino mass matrix expressed in the basis $%
\left( \nu _{L},\nu _{R}^{C},N_{R}^{C}\right) $, has the following
structure: 
\begin{equation}
M_{\nu }=\left( 
\begin{array}{ccc}
0_{3\times 3} & \varepsilon & m \\ 
\varepsilon ^{T} & 0_{2\times 2} & M \\ 
m^{T} & M^{T} & 0_{2\times 2}%
\end{array}
\right) ,  \label{Mnufull}
\end{equation}
where 
$\nu _{iL}$ ($i=1,2,3$) are active neutrinos, whereas $\nu _{kR}$ and $%
N_{kR} $ ($k=1,2$) are the sterile neutrinos. Their lepton numbers are $%
L(\nu_L)=L(\nu_R) = - L(N_R) =1$. Therefore, the only source of lepton
number violation is $m$-entry. 
For the linear seesaw mechanism to work properly, the entries of the full
neutrino mass matrix (\ref{Mnufull}) should obey the hierarchy $\varepsilon
_{in}<<m_{in}<<M_{np}$ ($i=1,2,3 $, $n,p=1,2$). 
In what follows we will discuss the models where the smallness of the
submatrix $\varepsilon$ is due to symmetries allowing its generation only at
one- and two-loop levels. 

We first explain how the tree level type I and tree level linear seesaw
mechanisms are forbidden in our models. 
This is done by precluding certain operators with ad hoc symmetries imposed on the models. 
To forbid the tree level type~I seesaw mechanism, we
look for a symmetry forbidding the operators:

\begin{equation}
\overline{l}_{iL}\widetilde{\phi }\nu _{nR},\hspace{1cm}m_{\nu _{R}}%
\overline{\nu }_{mR}\nu _{nR}^{C}
\end{equation}
whereas in the case of 
linear seesaw mechanism it is sufficient to ban only one
of the following operators:
\begin{equation}
\overline{l}_{iL}\widetilde{\phi }\nu _{nR},\hspace{1cm}\overline{l}_{iL}%
\widetilde{\phi }N_{nR},\hspace{1cm}m_{\nu _{R}}\overline{\nu }%
_{mR}N_{nR}^{C}\,.
\end{equation}
Here $\phi $ is the SM Higgs doublet, $l_{iL}$ ($i=1,2,3$) stand for the SM
leptonic doublets, while $\nu _{nR}$ and $N_{nR}$ ($n=1,2,3$) are gauge
singlets right handed Majorana neutrinos.
%
%
We choose to ban the $\overline{l}_{iL}%
\widetilde{\phi }\nu _{nR}$ Yukawa operator, whereas allowing the $\overline{l}_{iL}\widetilde{\phi }N_{nR}$ and $m_{\nu _{R}}\overline{\nu }_{mR}N_{nR}^{C}$ operators. To implement the above specified conditions we 
impose on the model a $Z_{4}$ discrete
symmetry, which is assumed to be spontaneously broken down to a preserved $\widetilde{Z}_{2}$ discrete symmetry.  This guarantees the radiative
nature of the linear seesaw mechanism, where
the submatrix $\varepsilon $ arises at one loop level. 
%
In this approach we also need to extend the SM scalar sector by
adding an inert $SU\left(2\right) _{L}$ scalar doublet $\eta$, two pairs of
electrically charged scalar singlets $S_{1}^{\pm }$, $S_{2}^{\pm }$, two
electrically neutral scalar singlets $\chi $, $\xi $ and three $SU\left(
2\right) _{L}$  singlet charged vector-like leptons $E_{i}$ ($i=1,2,3$).
With these fields we can built the following operators needed for the
construction of the submatrix $\varepsilon $ at one loop level:
\begin{equation}
\overline{l}_{iL}\eta E_{jR},\hspace{1cm}\overline{E}_{iL}S_{1}^{-}\nu _{nR},%
\hspace{1cm}\left( M_{E}\right) _{ij}\overline{E}_{iL}E_{jR},\hspace{1cm}%
S_{1}^{+}S_{2}^{-}\chi \xi ,\hspace{1cm}\varepsilon _{ab}\left( \eta ^{\ast
}\right) ^{a}\left( \phi ^{\ast }\right) ^{b}S_{2}^{+}\,.
\end{equation}
For generating the submatrices $m$ and $M$, one has to introduce the
operators:
\begin{equation}
\overline{l}_{iL}\widetilde{\phi }N_{nR},\hspace{1cm}M_{mn}\overline{\nu }%
_{mR}N_{nR}^{C}
\end{equation}

Let us note, that 
several of the above-introduced fields

play an important role in
the one loop level radiative seesaw mechanism that generates the SM charged
fermion masses. 

%
For this, it is additionally necessary to introduce an electrically neutral scalar singlet $\rho$ and three charged exotic leptons $E_{i}$ ($i=1,2,3$), as well as right-handed Majorana neutrinos $N_{nR } $ and $\nu _{nR}$ ($n=1,2$) in the singlet $SU\left( 2\right)$ representations.
The charged exotic leptons $E_{i}$ ($i=1,2,3$) mediate a one loop level radiative seesaw mechanism that gives rise to the SM charged lepton masses and to the neutrino mass submatrix $\varepsilon $. 
This mechanism can be implemented via the following operators:
\begin{equation}
\overline{l}_{iL}\eta E_{jR},\hspace{1cm}\overline{E}_{iL}\rho l_{jR},%
\hspace{1cm}\eta ^{\dag }\phi \rho ,\hspace{1cm}\left( \rho ^{\ast }\right)
^{2}\chi \xi ,\hspace{1cm}\left(M_{E}\right) _{ij}\overline{E}_{iL}E_{jR},
\end{equation}
To guarantee the radiative nature of the above described seesaw mechanisms, we need to 
add the spontaneously broken $%
Z_{2}$ and $Z_{4}$ discrete symmetries, with $Z_{4}$ broken down to a
residual $\widetilde{Z}_{2}$ discrete symmetry preserved at low energies.
Furthermore, the inclusion of the $Z_{4}$ symmetry will be crucial to forbid
the Majorana mass terms $M_{mn}^{\left( \nu \right) }\overline{\nu }_{mR}\nu
_{nR}^{C}$ and $M_{mn}^{\left( N\right) }\overline{N}_{mR}N_{nR}^{C}$ , thus
allowing us to have zero $2\times 2$ submatrices in Eq.~(\ref{Mnufull}). We
assume that the inert $SU\left( 2\right) $ doublet $\eta $ and the singlet
scalar $\rho $ have complex $Z_{4}$ charges, thus implying that they will be
charged under the residual $\widetilde{Z}_{2}$ symmetry. The inclusions of
these inert fields is necessary to radiatively generate the SM charged
lepton mass matrix at one loop level. To close the corresponding loop, the
scalar singlets $\chi $ and $\xi $, having real $Z_{4}$ charges, are required
in the scalar spectrum. Besides that, the radiative seesaw mechanism that
generates the submatrix $\varepsilon $ at one loop level is mediated by the
above mentioned inert doublet $\eta $, as well as by the electrically
charged scalar singlets $S_{1}^{\pm }$ and $S_{2}^{\pm }$, whose inclusion
is necessary for the implementation of this mechanism.

Now we proceed to discuss the case where the submatrix $\varepsilon $ is
generated at two loop level. 

This requires introduction of
six charged exotic leptons $E_{i}$, $\widetilde{E}_{i}$ ($i=1,2,3$), 
right handed Majorana neutrinos $N_{nR}$
and $\nu _{nR}$ ($n=1,2$) in the singlet representations of $SU\left(
2\right)$ 
as well as inert $SU\left( 2\right)$ doublet scalar $\eta$,
and gauge singlet scalars $\chi $, $\xi $, $\sigma $, $S_{1}^{\pm
}$, $S_{2}^{\pm }$. 
Generation of the submatrix $\varepsilon $ at two loop
level requires the inclusion of the following operators: 
\begin{equation}
\overline{l}_{iL}\eta E_{jR},\hspace{0.35cm}\overline{E}_{iL}S_{1}^{-}\nu
_{nR},\hspace{0.35cm}\left( M_{E}\right) _{ij}\overline{E}_{iL}E_{jR},%
\hspace{0.35cm}\overline{E}_{iL}\sigma \widetilde{E}_{jR},\hspace{0.35cm}%
\left( M_{\widetilde{E}}\right) _{ij}\overline{\widetilde{E}}_{iL}\widetilde{%
E}_{jR},\hspace{0.35cm}S_{1}^{+}S_{2}^{-}\chi \xi ,\hspace{0.35cm}%
\varepsilon _{ab}\left( \eta ^{\ast }\right) ^{a}\left( \phi ^{\ast }\right)
^{b}S_{2}^{+}\sigma ^{\ast }
\end{equation}%
The condition that the submatrix $\varepsilon$ appears only at the two-loop level requires introduction of the conserved discrete 
$Z_{2}^{\left( 2\right) }$ symmetry as well as of the spontaneously broken $Z_{2}^{\left( 1\right) }$ and $Z_{4}$ discrete symmetries, with $Z_{4}$ assumed to be broken down to a preserved $\widetilde{Z}_{2}$ discrete symmetry. 
%
In comparison
with the previously discussed one-loop realization,
here we need an extra
preserved discrete $Z_{2}^{\left( 2\right) }$ symmetry 
in order to prevent one-loop
radiative generation of the submatrix $\varepsilon $. 
%
Note that in this realization of the  two-loop linear seesaw mechanism, the masses of charged SM fermions are generated in exactly the same way as in the previously described one-loop case.

\subsection{Model 1: One-loop radiative linear seesaw model}

\label{modelA} We consider an extended Inert Doublet Model (IDM), where the
scalar content is enlarged by the inclusion of several gauge singlet
scalars, whereas the lepton sector is augmented by considering right handed
Majorana neutrinos and charged exotic vector like leptons. The SM gauge
symmetry is extended by the inclusion of the spontaneously broken $%
Z_{2}\times Z_{4}$ discrete symmetry. The $Z_{4}$ discrete symmetry is
broken down to a preserved $\widetilde{Z}_{2}$ discrete symmetry, which is a
stabilizer of Dark Matter (DM) particle candidate. Schematically, the
symmetry breaking chain goes as follows, 
\begin{eqnarray}
\mathbb{G}_1&=&\hspace{5mm}SU(3)_{C}\times SU\left( 2\right) _{L}\times
U\left( 1\right) _{Y}\times Z_2\times Z_4  \notag \\
&&\hspace{35mm}\Downarrow v_{\xi},v_{\chi}  \notag \\
&&\hspace{10mm}SU(3)_{C}\times SU\left( 2\right) _{L}\times U\left(
1\right)_{Y}\times \widetilde{Z}_{2}  \notag \\
&&\hspace{35mm}\Downarrow v  \notag \\
&&\hspace{15mm}SU(3)_{C}\times U\left( 1\right)_\text{em}\times\widetilde{Z}%
_{2}
\end{eqnarray}
where $\Tilde{Z}_2 \subset Z_4$. The SM Higgs VEV we denoted as $%
\langle\phi^0\rangle=v/\sqrt{2}$. The global $Z_2\times Z_4$ discrete
symmetry is spontaneously broken at the TeV scale by the vacuum expectation
values (VEVs) of the gauge singlet scalars $\langle\xi\rangle=v_\xi$, $%
\langle\chi\rangle=v_\chi$. %

The scalar and leptonic fields with their assignments under the model
symmetry group $\mathbb{G}_1$ are shown in Tables \ref{scalars} and \ref%
{leptons}, respectively. 
This defines the leptonic Yukawa interaction in this model 
\begin{eqnarray}  \label{eq:1-L-Lagr}
-\mathcal{L}_{Y}^{\left( l\right) }
&=&\sum_{i=1}^{3}\sum_{j=1}^{3}y_{ij}^{\left( E\right) }\overline{l}%
_{iL}\eta E_{jR}+\sum_{i=1}^{3}\sum_{j=1}^{3}x_{ij}^{\left( l\right) }%
\overline{E}_{iL}\rho l_{jR}+\sum_{i=1}^{3}\sum_{n=1}^{2}x_{in}^{\left( \nu
\right) }\overline{E}_{iL}S_{1}^{-}\nu
_{nR}+\sum_{i=1}^{3}\sum_{n=1}^{2}y_{in}^{\left( N\right) }\overline{l}_{iL}%
\widetilde{\phi }N_{nR} \\
&&+\sum_{m=1}^{2}\sum_{n=1}^{2}M_{mn}\overline{\nu }_{mR}N_{nR}^{C}+%
\sum_{i=1}^{3}\sum_{j=1}^{3}\left( M_{E}\right) _{ij}\overline{E}
_{iL}E_{jR}+h.c  \notag
\end{eqnarray}


Note that the residual $\widetilde{Z}_{2}$ discrete symmetry is preserved at
low energies since we require that the $SU\left( 2\right) _{L}$ doublet
scalar $\eta$ and the scalar singlets $\rho$ do not develop vacuum
expectation values. As we will show in subsequent sections a viable DM
candidate is the lightest among the $\widetilde{Z}_{2}$-odd scalar fields $%
\func{Re}\eta ^{0}$, $\func{Im}\eta ^{0}$, $\func{Re}\rho $, $\func{Im}\rho$.

%
Furthermore, the electrically charged components of the $SU\left( 2\right)
_{L}$ doublet scalar $\eta $, together with the electrically charged gauge
singlet scalars $S_{1}^{-}$, $S_{2}^{-}$\ \ and the vector like leptons $%
E_{i}$ ($i=1,2,3$) will induce a one loop level radiative seesaw mechanism
that generates the Dirac neutrino mass matrix, as shown in the Feynman
diagram of Figure \ref{Neutrino1loopdiagram}. We further include right
handed Majorana neutrinos $\nu _{nR}$,$\ N_{nR}$ ($n=1,2$) (having opposite $%
Z_{4}$ charges), in order to implement a one loop level radiative linear
seesaw mechanism that produces the tiny masses of the light active
neutrinos. Furthermore, the heavy vector like leptons $E_{i}$ ($i=1,2,3$),
induce a radiative seesaw mechanism at one loop level that generates the SM
charged lepton masses. It is worth mentiong that having three vector like
leptons $E_{i}$ ($i=1,2,3$), singlets under $SU\left( 2\right) _{L}$, is the
minimal amount of charged exotic leptons needed to generate the SM charged
lepton masses. Furthermore, the extra $Z_{2}\times Z_{4}$ discrete symmetry
selects the allowed entries of the full neutrino mass matrix, thus allowing
a successful implementation of the linear seesaw mechanism that produces the
light active neutrino masses.

\begin{table}[tp]
\begin{tabular}{|c|c|c|c|c|c|c|}
\hline
& $SU\left( 3\right)_{C}$ & $SU\left( 2\right)_{L}$ & $U\left( 1\right)_{Y}$
& $Z_{2}$ & $Z_{4}$ & $\tilde{Z}_2$ \\ \hline
$\phi$ & $1$ & $2$ & $\frac{1}{2}$ & $1$ & $1$ & $1$ \\ \hline
$\eta$ & $1$ & $2$ & $\frac{1}{2}$ & $1$ & $i$ & $-1$ \\ \hline
$\rho$ & $1$ & $1$ & $0$ & $1$ & $i$ & $-1$ \\ \hline
$\xi$ & $1$ & $1$ & $0$ & $-1$ & $-1$ & $1$ \\ \hline
$\chi$ & $1$ & $1$ & 0 & $-1$ & $1$ & $1$ \\ \hline
$S_{1}^{\pm}$ & $1$ & $1$ & $\pm 1$ & $1$ & $\mp i$ & $-1$ \\ \hline
$S_{2}^{\pm}$ & $1$ & $1$ & $\pm 1$ & $1$ & $\pm i$ & $-1$ \\ \hline
\end{tabular}%
\caption{Model 1. Scalar assignments under $\mathbb{G}_1=SU\left(
3\right)_{C}\times SU\left( 2\right)_{L}\times U\left( 1\right)_{Y}\times
Z_2\times Z_4$ symmetry in the one loop radiative linear seesaw model.}
\label{scalars}
\end{table}

\begin{table}[tp]
\begin{tabular}{|c|c|c|c|c|c|c|}
\hline
& $SU\left( 3\right)_{C}$ & $SU\left( 2\right)_{L}$ & $U\left( 1\right)_{Y}$
& $Z_{2}$ & $Z_{4}$ & $\tilde{Z}_2$ \\ \hline
$l_{iL}$ & $1$ & $2$ & $-\frac{1}{2}$ & $1$ & $-i$ & $-1$ \\ \hline
$l_{iR}$ & $1$ & $1$ & $-1$ & $1$ & $i$ & $-1$ \\ \hline
$\nu _{nR}$ & $1$ & $1$ & $0$ & $1$ & $i$ & $-1$ \\ \hline
$N_{nR}$ & $1$ & $1$ & $0$ & $1$ & $-i$ & $-1$ \\ \hline
$E_{iL}$ & $1$ & $1$ & $-1$ & $1$ & $-1$ & $1$ \\ \hline
$E_{iR}$ & $1$ & $1$ & $-1$ & $1$ & $-1$ & $1$ \\ \hline
\end{tabular}%
\caption{Model 1. Lepton assigments under $\mathbb{G}_1=SU\left( 3\right)
_{C}\times SU\left( 2\right) _{L}\times U\left( 1\right) _{Y}\times
Z_{2}\times Z_{4}$ symmetry in the one loop radiative linear seesaw model. }
\label{leptons}
\end{table}



\subsection{Model 2: Two-loop radiative linear seesaw model}

\label{modelB} Now we consider an extension of the previously described one
loop radiative linear seesaw model, where the masses of the light active
neutrinos appear at two loop level. The scalar and leptonic content will be
similar to the previous model. In addition to the fields considered there,
we introduce charged exotic vector like leptons $\widetilde{E}_{i}$ ($%
i=1,2,3 $)\ as well as the electrically neutral gauge singlet scalar $\sigma 
$, assumed to be charged under a preserved $Z_{2}^{\left( 2\right) }$
symmetry.

The SM gauge symmetry is supplemented by the inclusion of the $Z_{2}^{\left(
1\right) }\times Z_{2}^{\left( 2\right) }\times Z_{4}$ discrete group. 
%
The full symmetry $\mathbb{G}$ of the model exhibits the following breaking
scheme 
\begin{eqnarray}
\mathbb{G}_2 &=&\hspace{5mm}SU(3)_{C}\times SU\left( 2\right) _{L}\times
U\left( 1\right) _{Y}\times Z_{2}^{\left( 1\right) }\times Z_{2}^{\left(
2\right) }\times Z_{4} \\
&&\hspace{35mm}\Downarrow v_{\xi},v_{\chi}  \notag \\
&&\hspace{15mm}SU(3)_{C}\times SU\left( 2\right) _{L}\times U\left(
1\right)_{Y}\times Z_{2}^{\left( 2\right) }\times\widetilde{Z}_{2}  \notag \\
&&\hspace{35mm}\Downarrow v  \notag \\
&&\hspace{15mm}SU(3)_{C}\times U\left( 1\right)_\text{em}\times
Z_{2}^{\left( 2\right) }\times\widetilde{Z}_{2}
\end{eqnarray}
where $\Tilde{Z}_2 \subset Z_4$. The scalar and leptonic fields and their
assignments under the model symmetry group $\mathbb{G}$ 
are shown in Tables~\ref{leptons2} and \ref{scalars2}, respectively. With
this field content and symmetries we have the following relevant leptonic
Yukawa interactions 
\begin{eqnarray}  \label{eq:2-L-Lagr}
-\mathcal{L}_{Y}^{\left( l\right) }
&=&\sum_{i=1}^{3}\sum_{j=1}^{3}y_{ij}^{\left( E\right) }\overline{l}%
_{iL}\eta E_{jR}+\sum_{i=1}^{3}\sum_{j=1}^{3}x_{ij}^{\left( l\right) }%
\overline{E}_{iL}\rho l_{jR}+\sum_{i=1}^{3}\sum_{n=1}^{2}x_{in}^{\left( \nu
\right) }\overline{\widetilde{E}}_{iL}S_{1}^{-}\nu
_{nR}+\sum_{i=1}^{3}\sum_{n=1}^{2}y_{in}^{\left( N\right) }\overline{l}_{iL}%
\widetilde{\phi }N_{nR} \\
&&+\sum_{i=1}^{3}\sum_{j=1}^{3}x_{ij}^{\left( E\right) }\overline{E}%
_{iL}\sigma \widetilde{E}_{jR}+\sum_{m=1}^{2}\sum_{n=1}^{2}M_{mn}\overline{%
\nu }_{mR}N_{nR}^{C}+\sum_{i=1}^{3}\sum_{j=1}^{3}\left( M_{E}\right) _{ij}%
\overline{E}_{iL}E_{jR}+\sum_{i=1}^{3}\sum_{j=1}^{3}\left( M_{\widetilde{E}%
}\right) _{ij}\overline{\widetilde{E}}_{iL}\widetilde{E}_{jR}+h.c  \notag
\end{eqnarray}

In this model $Z_{2}^{\left( 2\right) }\times \widetilde{Z}_{2}$ discrete
symmetry is preserved at low energies since we require that the $SU\left(
2\right) _{L}$ doublet scalar $\eta $ as well as the electrically neutral
scalar singlets $\rho $ and $\sigma $ do not acquire vacuum expectation
values.

As we will show in subsequent sections a viable DM candidate is the %
%
lightest among the $\widetilde{Z}_{2}$-odd scalar fields $\func{Re}\eta ^{0}$%
, $\func{Im}\eta ^{0}$, $\func{Re}\rho $, $\func{Im}\rho $, $\func{Re}\sigma 
$, $\func{Im}\sigma$. 

\begin{table}[tp]
\begin{tabular}{|c|c|c|c|c|c|c|c|}
\hline
& $SU\left( 3\right)_{C}$ & $SU\left( 2\right)_{L}$ & $U\left( 1\right)_{Y}$
& $Z_{2}^{\left( 1\right)}$ & $Z_{2}^{\left(2\right)}$ & $Z_{4}$ & $\tilde{Z}%
_2$ \\ \hline
$l_{iL}$ & $1$ & $2$ & $-\frac{1}{2}$ & $1$ & $1$ & $-i$ & $-1$ \\ \hline
$l_{iR}$ & $1$ & $1$ & $-1$ & $1$ & $1$ & $i$ & $-1$ \\ \hline
$\nu _{nR}$ & $1$ & $1$ & $0$ & $1$ & $1$ & $i$ & $-1$ \\ \hline
$N_{nR}$ & $1$ & $1$ & $0$ & $1$ & $1$ & $-i$ & $-1$ \\ \hline
$E_{iL}$ & $1$ & $1$ & $-1$ & $1$ & $1$ & $-1$ & $1$ \\ \hline
$E_{iR}$ & $1$ & $1$ & $-1$ & $1$ & $1$ & $-1$ & $1$ \\ \hline
$\widetilde{E}_{iL}$ & $1$ & $1$ & $-1$ & $1$ & $-1$ & $-1$ & $1$ \\ \hline
$\widetilde{E}_{iR}$ & $1$ & $1$ & $-1$ & $1$ & $-1$ & $-1$ & $1$ \\ \hline
\end{tabular}%
\caption{Model 2. Lepton assigments under $\mathbb{G}_2=SU\left( 3\right)
_{C}\times SU\left( 2\right) _{L}\times U\left( 1\right) _{Y}\times
Z_{2}^{\left( 1\right) }\times Z_{2}^{\left( 2\right) }\times Z_{4}$
symmetry in the two loop radiative linear seesaw model. }
\label{leptons2}
\end{table}

\begin{table}[tp]
\begin{tabular}{|c|c|c|c|c|c|c|c|}
\hline
& $SU\left( 3\right) _{C}$ & $SU\left( 2\right) _{L}$ & $U\left( 1\right)
_{Y}$ & $Z_{2}^{\left( 1\right) }$ & $Z_{2}^{\left( 2\right) }$ & $Z_{4}$ & $%
\tilde{Z}_{2}$ \\ \hline
$\phi $ & $1$ & $2$ & $\frac{1}{2}$ & $1$ & $1$ & $1$ & $1$ \\ \hline
$\eta $ & $1$ & $2$ & $\frac{1}{2}$ & $1$ & $1$ & $i$ & $-1$ \\ \hline
$\sigma $ & $1$ & $1$ & $0$ & $1$ & $-1$ & $1$ & $1$ \\ \hline
$\rho $ & $1$ & $1$ & $0$ & $1$ & $1$ & $i$ & $-1$ \\ \hline
$\xi $ & $1$ & $1$ & $0$ & $-1$ & $1$ & $-1$ & $1$ \\ \hline
$\chi $ & $1$ & $1$ & $0$ & $-1$ & $1$ & $1$ & $1$ \\ \hline
$S_{1}^{\pm }$ & $1$ & $1$ & $\pm 1$ & $1$ & $-1$ & $\mp i$ & $-1$ \\ \hline
$S_{2}^{\pm }$ & $1$ & $1$ & $\pm 1$ & $1$ & $-1$ & $\pm i$ & $-1$ \\ \hline
\end{tabular}%
\caption{Model 2. Scalar assignments under $\mathbb{G}_{2}=SU\left( 3\right)
_{C}\times SU\left( 2\right) _{L}\times U\left( 1\right) _{Y}\times
Z_{2}^{\left( 1\right) }\times Z_{2}^{\left( 2\right) }\times Z_{4}$
symmetry in the two loop radiative linear seesaw model. }
\label{scalars2}
\end{table}
\begin{figure}[th]
\includegraphics[width=0.9\textwidth]{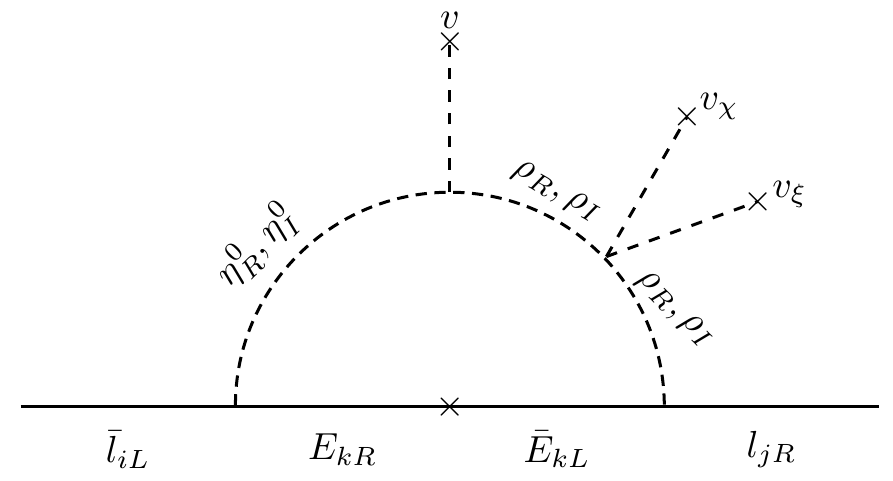}
\caption{Models 1 and 2. One-loop Feynman diagram contributing to the
charged lepton mass matrix. Here $i,j,k=1,2,3$.}
\label{Chargedleptonloopdiagram}
\end{figure}


\section{Scalar Potential}

In this section we analyze the scalar potential of model 1. The scalar
potential invariant under the symmetries of the model 
has the form: 
\begin{align}
V=& -\mu _{\phi }^{2}(\phi ^{\dagger }\phi )-\mu _{\eta }^{2}(\eta ^{\dagger
}\eta )-\mu _{\rho }^{2}(\rho ^{\dagger }\rho )-\mu _{\xi }^{2}(\xi
^{\dagger }\xi )-\mu _{\chi }^{2}(\chi ^{\dagger }\chi )+\lambda _{1}(\phi
^{\dagger }\phi )(\phi ^{\dagger }\phi )+\lambda _{2}(\eta ^{\dagger }\eta
)(\eta ^{\dagger }\eta )  \notag \\
& +\lambda _{3}(\rho ^{\dagger }\rho )(\rho ^{\dagger }\rho )+\lambda
_{4}(\xi ^{\dagger }\xi )(\xi ^{\dagger }\xi )+\lambda _{5}(\chi ^{\dagger
}\chi )(\chi ^{\dagger }\chi )+\lambda _{6}(\phi ^{\dagger }\eta )(\eta
^{\dagger }\phi )+\lambda _{7}(\eta ^{\dagger }\rho )(\rho ^{\dagger }\eta
)+\lambda _{8}(\eta ^{\dagger }\eta )(\rho ^{\dagger }\rho )  \notag \\
& +\lambda _{9}(\phi ^{\dagger }\xi )(\xi ^{\dagger }\phi )+\lambda
_{10}(\phi ^{\dagger }\chi )(\chi ^{\dagger }\phi )+\lambda _{11}(\phi
^{\dagger }\phi )(\xi ^{\dagger }\xi )+\lambda _{12}(\phi ^{\dagger }\phi
)(\chi ^{\dagger }\chi )+\lambda _{13}(\xi ^{\dagger }\xi )(\chi ^{\dagger
}\chi )  \notag \\
& +\lambda _{14}(\xi ^{\dagger }\chi )(\chi ^{\dagger }\xi )+\lambda
_{15}(\phi ^{\dagger }\phi )(\eta ^{\dagger }\eta )+\lambda _{16}(\phi
^{\dagger }\phi )(\rho ^{\dagger }\rho )+f(\eta ^{\dagger }\phi \rho +\text{%
h.c})+\lambda _{17}(\eta ^{\dagger }\chi )(\chi ^{\dagger }\eta )  \notag \\
& +\lambda _{18}(\rho ^{\dagger }\chi )(\chi ^{\dagger }\rho )+\lambda
_{19}(\eta ^{\dagger }\xi )(\xi ^{\dagger }\eta )+\lambda _{20}(\rho
^{\dagger }\xi )(\xi ^{\dagger }\rho )+\lambda _{21}(\rho ^{2}\xi \chi +%
\text{h.c})  \label{eq:scalar_potential}
\end{align}
%
where the $SU(2)_{L}$ scalar doublets are given by 
\begin{align}
\phi = 
\begin{pmatrix}
\phi^{\pm} \\ 
\phi^0%
\end{pmatrix}%
, \ \ \ \eta = 
\begin{pmatrix}
\eta^{\pm} \\ 
\eta^0%
\end{pmatrix}%
\end{align}

Here, $\phi$ is the SM Higgs doublet and $\eta$ is the inert $SU(2)$ scalar
doublet. The coupling terms of $\phi S_2 \eta$, $\eta \phi \sigma S_2$ and $%
S_2 S_1 \chi \xi$ are considered in the vertices of diagrams in Figure \ref%
{Neutrino1loopdiagram} and Figure \ref{Neutrino2loopdiagram} whereas in the
scalar potential in Eq.~(\ref{eq:scalar_potential}) we consider the common
terms in Model 1 and Model 2 to study phenomenological aspects like muon g-2
in Section \ref{sec:G-2} and dark matter in Section \ref{section-DM}.

For finding constraints on the parameters $\lambda_i$ from the condition
that the scalar potential $V$ is bounded from below, we just need to examine
the quartic terms of the scalar potential as in \cite{BHATTACHARYYA_2016}.
Here we considered quartic terms in the scalar potential that are relevant
for $g-2$ and Dark Matter phenomenology. For convenience we define $%
a=\phi^\dagger \phi$, $b=\eta^\dagger \eta$, $c=\rho^\dagger \rho$, $d= \Re
(\phi^\dagger \eta)$, $e= \Im (\phi^\dagger \eta)$, $g=\Re (\eta^\dagger
\rho)$, $h=\Im (\eta^\dagger \rho)$, $z=\eta^{\dagger}\phi \rho$, $j=\Re
(\eta^{\dagger}\chi)$, $k=\Im (\eta^{\dagger}\chi)$, $l=\Re
(\rho^{\dagger}\chi)$, $m=\Im (\rho^{\dagger}\chi)$, $n=\Re(\eta^{\dagger}%
\xi)$, $p=\Im(\eta^{\dagger}\xi)$, $q=\Re(\rho^\dagger \xi)$, $%
r=\Im(\rho^\dagger \xi)$, $x=\rho^2\xi\chi$. With this notation we can write
the quartic terms $V_4\subset V$ as 

\begin{align*}
V_4 =& \left( \lambda_1^{1/2}a - \lambda_2^{1/2}b \right)^2 + \left(
\lambda_2^{1/2}b - \lambda_3^{1/2}c \right)^2 + \left(\lambda_{15} +
2\lambda_1^{1/2}\lambda_2^{1/2}\right)\left(ab-d^2-e^2\right) \\
&+ (\lambda_8 + 2(\lambda_2 \lambda_3)^{1/2})(bc-g^2-h^2) +
\left(\lambda_{8} + \lambda_{7} + 2\left(\lambda_2 \lambda_3\right)^{1/2}
\right) \left(g^2+h^2\right) \\
& + \left(\lambda_{15} + \lambda_{6} + 2\left(\lambda_1
\lambda_2\right)^{1/2} \right) \left(d^2+e^2\right) +\lambda_{16}(ac) + f z
+\lambda_{17} (j^2 + k^2) \\
& + \lambda_{18} (l^2 + m^2) + \lambda_{19} (n^2 + p^2) + \lambda_{20} (q^2
+ r^2) + \lambda_{21} (x + \text{h.c.})
\end{align*}

We require that there are no directions in the field space along which $V
\to -\infty$. This leads to the following constraints on the $\lambda_i$: 
%

\begin{itemize}
\item If $a=0$ then $d=e=z=0$, so we obtain $\lambda_2 > 0$

\item If $b=0$ we obtain $\lambda_1 > 0$ and $\lambda_3 > 0$

\item If $c=0$ and $ab = d^2 + e^2$ we obtain $\lambda_{15} + 2\sqrt{%
\lambda_1 \lambda_2} >0 $

\item If $a=\sqrt{\lambda_2/\lambda_1}b$ and $g=h=d=e=0$ then $\lambda_{8} +
2\sqrt{\lambda_2 \lambda_3} >0 $

\item If $b=\sqrt{\lambda_3/\lambda_2}c$, $bc=g^2 + h^2$ and $d=e=0$ then $%
\lambda_{8} +\lambda_{7}+ 2\sqrt{\lambda_2 \lambda_3} >0 $

\item $\lambda_{15} +\lambda_{6}+ 2\sqrt{\lambda_1 \lambda_2} >0 $
\end{itemize}

These constraints are used in our numerical analysis of muon anomalous
magnetic moment and dark matter.

Due to the scalar field charge assignments the mass matrix $M^2_{\eta \rho}$
is the same in both our Models 1 and 2, described in the previous sections.
In the basis $\func{Re}\eta ^{0}$, $\func{Re}\rho $, $\func{Im}\eta ^{0}$, $%
\func{Im}\rho$ 
this matrix is given by: 
\begin{align}
M^2_{\eta \rho} = \left( 
\begin{array}{cccc}
m_{11} & m_{12} & 0 & 0 \\ 
m_{21} & m_{22} & 0 & 0 \\ 
0 & 0 & m_{33} & m_{34} \\ 
0 & 0 & m_{43} & m_{44}%
\end{array}
\right)  \label{Metarho}
\end{align}
where the matrix entries are:
\begin{align}
\label{eq:M-eta-rho-1}
m_{11} &=-\mu ^2{}_{\eta }+\frac{1}{2} \lambda _{19} v_{\xi }^2+\frac{1}{2}
\lambda _{17} v_{\chi }^2+\frac{1}{2} \lambda _6 v_{\phi }^2+\frac{1}{2}
\lambda _{15} v_{\phi }^2, \\
m_{22} &= -\mu ^2{}_{\rho }+\lambda _{21} v_{\xi } v_{\chi }+\frac{1}{2}
\lambda _{20} v_{\xi }^2+\frac{1}{2} \lambda _{18} v_{\chi }^2+\frac{1}{2}
\lambda _{16} v_{\phi }^2, \\
m_{33} &= -\mu ^2{}_{\eta }+\frac{1}{2} \lambda _{19} v_{\xi }^2+\frac{1}{2}
\lambda _{17} v_{\chi }^2+\frac{1}{2} \lambda _6 v_{\phi }^2+\frac{1}{2}
\lambda _{15} v_{\phi }^2, \\
m_{44} &= -\mu ^2{}_{\rho }-\lambda _{21} v_{\xi } v_{\chi }+\frac{1}{2}
\lambda _{20} v_{\xi }^2+\frac{1}{2} \lambda _{18} v_{\chi }^2+\frac{1}{2}
\lambda _{16} v_{\phi }^2, \\
m_{12} &= m_{21} = \frac{f v_{\phi }}{\sqrt{2}}, \\
\label{eq:M-eta-rho-6}
m_{34} &= m_{43} = -\frac{f v_{\phi }}{\sqrt{2}}\,.
\end{align}
The upper left and lower right blocks of the matrix given in Eq.~(\ref%
{Metarho}) correspond to the squared scalar mass matrices for the dark CP
even and CP odd scalars, respectively. Their diagonalization yields the
following physical CP even and CP odd mass eigenstates defined as follows: 
\begin{equation}
\left( 
\begin{array}{c}
H_{1} \\ 
H_{2}%
\end{array}%
\right) =\left( 
\begin{array}{cc}
\cos \theta _{H} & \sin \theta _{H} \\ 
-\sin \theta _{H} & \cos \theta _{H}%
\end{array}%
\right) \left( 
\begin{array}{c}
\eta _{R} \\ 
\rho _{R}%
\end{array}%
\right) ,\hspace{1cm}\left( 
\begin{array}{c}
A_{1} \\ 
A_{2}%
\end{array}%
\right) =\left( 
\begin{array}{cc}
\cos \theta _{A} & \sin \theta _{A} \\ 
-\sin \theta _{A} & \cos \theta _{A}%
\end{array}%
\right) \left( 
\begin{array}{c}
\eta _{I} \\ 
\rho _{I}%
\end{array}%
\right) .  \label{Darkscalars}
\end{equation}
{\color{blue} Here, $\eta_{R,I} = \func{Re}\eta^{0}, \func{Im}\eta^{0}$.} These relations will be used in the analysis of the lepton mass generation,
muon anomalous magnetic moment and dark matter we will carry out in the
following sections.

\section{Lepton sector masses}

Here we show that, both above specified one- and two-loop models offer
radiative mechanisms for the generation of charged lepton and neutrino
masses. In both models the charged lepton masses are generated at one-loop
level, according to Fig.~\ref{Chargedleptonloopdiagram}. On the other hand,
neutrino masses arise at one- and two-loop levels depending on the model, as
shown in Figs.~\ref{Neutrino1loopdiagram}, \ref{Neutrino2loopdiagram}.

\subsection{Charged lepton masses}

From the SM charged lepton Yukawa terms in (\ref{eq:1-L-Lagr}) and (\ref%
{eq:2-L-Lagr}), we find that the mass matrix for SM charged leptons in
models 1 and 2 is represented by the diagram in Fig.~\ref%
{Chargedleptonloopdiagram}. Analytically we have 
\begin{equation}
\left( M_{l}\right) _{ij}=\sum_{k=1}^{3}\frac{y_{ik}^{\left( E\right)
}x_{kj}^{\left( l\right) }m_{E_{k}}}{16\pi ^{2}}\left\{ \left[ F\left(
m_{H_{1}}^{2},m_{E_{k}}^{2}\right) -F\left(
m_{H_{2}}^{2},m_{E_{k}}^{2}\right) \right] \sin 2\theta _{H}-\left[ F\left(
m_{A_{1}}^{2},m_{E_{k}}^{2}\right) -F\left(
m_{A_{2}}^{2},m_{E_{k}}^{2}\right) \right] \sin 2\theta _{A}\right\} ,
\end{equation}
where $F\left(m_{1}^{2},m_{2}^{2}\right) $ is the function defined as, 
\begin{equation}
F\left(m_{1}^{2},m_{2}^{2}\right) =\frac{m_{1}^{2}}{m_{1}^{2}-m_{2}^{2}}\ln
\left( \frac{m_{1}^{2}}{m_{2}^{2}}\right) .
\end{equation}%
$m_{H_{1}}$ and $m_{H_{2}}$ are the masses of the physical CP even scalars,
whereas $m_{A_{1}}$ and $m_{A_{2}}$ are those of the inert pseudoscalars.

The SM charged lepton mass matrix can be parametrized as follows: 
\begin{eqnarray}
M_{l} &=&A_{l}J_{E}^{-1}B_{l}^{T},\hspace{0.7cm}\hspace{0.7cm}J_{E}=\left( 
\begin{array}{ccc}
\frac{1}{16\pi ^{2}}m_{E_{1}}K_{E}^{\left( 1\right) } & 0 & 0 \\ 
0 & \frac{1}{16\pi ^{2}}m_{E_{2}}K_{E}^{\left( 2\right) } & 0 \\ 
0 & 0 & \frac{1}{16\pi ^{2}}m_{E_{3}}K_{E}^{\left( 3\right) }%
\end{array}%
\right),
\end{eqnarray}
where: 
\begin{eqnarray}
K_{E}^{\left( n\right) } &=&\left[ F\left(
m_{H_{1}}^{2},m_{E_{n}}^{2}\right) -F\left(
m_{H_{2}}^{2},m_{E_{n}}^{2}\right) \right] \sin 2\theta _{H}-\left[ F\left(
m_{A_{1}}^{2},m_{E_{n}}^{2}\right) -F\left(
m_{A_{2}}^{2},m_{E_{n}}^{2}\right) \right] \sin 2\theta _{A} ,\hspace{0.5cm}%
n=1,2,3  \notag \\
A_{l}&=&V_{L}^{\left( l\right) }\widetilde{M}_{l}^{\frac{1}{2}}J_{E}^{\frac{1%
}{2}},\hspace{0.7cm}\hspace{0.7cm}B_{l}=V_{R}^{\left( l\right) }\widetilde{M}%
_{l}^{\frac{1}{2}}J_{E}^{\frac{1}{2}},\hspace{0.7cm}\hspace{0.7cm}\widetilde{%
M}_{l}=\left( 
\begin{array}{ccc}
m_{e} & 0 & 0 \\ 
0 & m_{\mu } & 0 \\ 
0 & 0 & m_{\tau }%
\end{array}%
\right)
\end{eqnarray}
Thus, both models have enough parametric freedom to successfully accommodate
the SM charged lepton masses. Despite the fact that the all SM charged
lepton masses arise at one loop level, the hierarchy between such masses can
be successfully accommodated by having some moderate hierarchy as well as a
deviation from the scenario of universality of the charged lepton Yukawa
couplings. It is worth mentioning that since the SM charged lepton mass
matrix arises at one loop level, the effective charged lepton Yukawa
couplings are proportional to a product of two other dimensionless
couplings, thus implying that a moderate hierarchy in those couplings can
give rise to a quadratically larger hierarchy in the effective couplings.

\subsection{Neutrino mass matrix}

\label{sec:NuMass} 
\begin{figure}[th]
\includegraphics[width=0.9\textwidth]{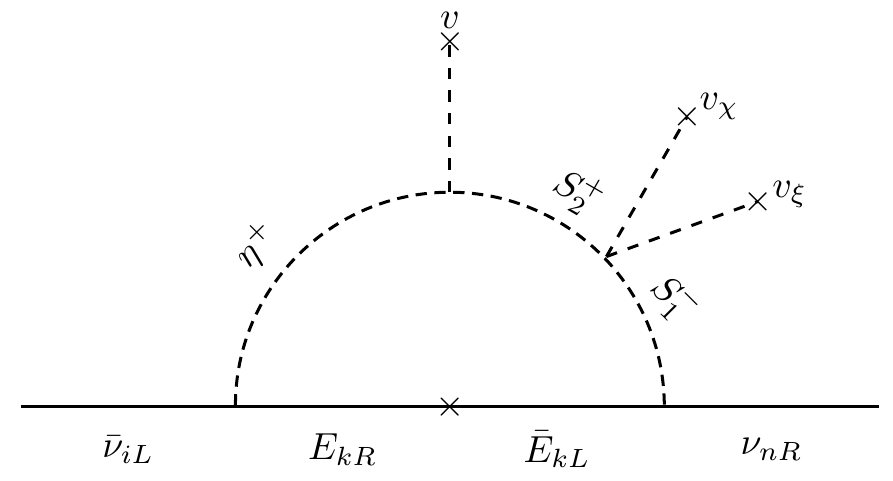}
\caption{Model 1. One-loop Feynman diagram contributing to the neutrino mass
submatrix $\protect\varepsilon$. Here $i,k,r=1,2,3$, $n=1,2$.}
\label{Neutrino1loopdiagram}
\end{figure}

\begin{figure}[th]
\includegraphics[width=0.9\textwidth]{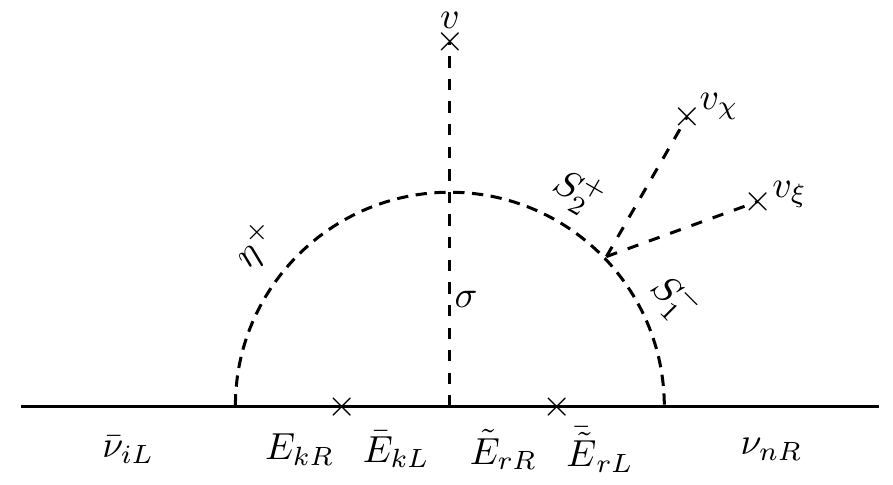}
\caption{Model 2. Two-loop Feynman diagram contributing to the neutrino mass
submatrix $\protect\varepsilon$. Here $i,k,r=1,2,3$, $n=1,2$.}
\label{Neutrino2loopdiagram}
\end{figure}

From the neutrino Yukawa interactions of both radiative
models described above, we find the following neutrino mass terms: 
\begin{equation}  \label{eq:Mass-Matr2}
-\mathcal{L}_{mass}^{\left( \nu \right) }=\frac{1}{2}\left( 
\begin{array}{ccc}
\overline{\nu _{L}^{C}} & \overline{\nu _{R}} & \overline{N_{R}}%
\end{array}%
\right) M_{\nu }\left( 
\begin{array}{c}
\nu _{L} \\ 
\nu _{R}^{C} \\ 
N_{R}^{C}%
\end{array}
\right) +H.c. = \varepsilon \, \overline{\nu_L} \nu_R + m\, \overline{\nu_L}
N_R + M\, \overline{\nu_R^C} N_R + H.c.
\end{equation}
where neutrino mass matrix $M_\nu$ is given by Eq.~(\ref{Mnufull}), after
which we also specified lepton number assignment of the neutrino sector. As
seen, the only source of lepton number violation is the $m$-term. In our
models this submatrix is generated after spontaneous breaking of the
electroweak symmetry by a VEV $\langle \phi^0\rangle = v$ of the SM Higgs
doublet $\phi$, which also breaks $U(1)_L$ of lepton number symmetry of both
models with the charge assignments $L(\nu_L)=L(\nu_R) = - L(N_R) = L(\phi)/2
= L(\eta)/2 = 1$. Therefore, as seen from Eqs. (\ref{eq:1-L-Lagr}) and (\ref%
{eq:2-L-Lagr}) we have 
\begin{equation}
m_{in}=y_{in}^{\left( N\right) }\frac{v}{\sqrt{2}},\hspace{0.7cm}\hspace{%
0.7cm}i=1,2,3,\hspace{0.7cm}\hspace{0.7cm}n=1,2.
\end{equation}
Since the global $U(1)_L$ is spontaneously broken by $\langle \phi^0\rangle$%
, one typically expects appearance of the corresponding SM-non-sterile
massless Majoron, which is phenomenologically unacceptable. Fortunately, it
does not appear as a physical state. The simultaneous spontaneous breaking
the SM gauge symmetry and $U(1)_L$ guaranties that the Majoron coincides
with the electrically neutral CP-odd would-be-Goldstone absorbed by $Z$%
-boson. The Dirac submatrix $\varepsilon$ is generated at one and two loops
in the models of sections \ref{modelA} and \ref{modelB}, respectively. The
corresponding Feynman diagram 
are shown in Figs.~\ref{Neutrino1loopdiagram} and \ref{Neutrino2loopdiagram}.

The submatrix $\varepsilon $ takes the form: 
\begin{equation}
\varepsilon _{in}\simeq \left\{ 
\begin{array}{l}
\dsum\limits_{j=1}^{3}\dsum\limits_{k=1}^{3}\frac{y_{ik}^{\left( E\right)
}x_{kn}^{\left( \nu \right) }m_{E_{k}}}{16\pi ^{2}}\left( R_{C}\right)
_{1j}\left( R_{C}\right) _{2j}F\left( m_{H_{j}^{\pm
}}^{2},m_{E_{k}}^{2}\right) ,\hspace{0.5cm}\mbox{for the one loop model} \\ 
\\ 
\dsum\limits_{k=1}^{2}\frac{\lambda vR_{1k}R_{2k}J\left( \frac{m_{\sigma
}^{2}}{m_{H_{k}^{\pm }}^{2}}\right)}{96\pi ^{2}m_{S}^{2}} \times \\ 
\times \left[ y^{\left( E\right) }M_{E}\left( x^{\left( E\right) }\right)
^{\dag }M_{\widetilde{E}}^{T}\left( x^{\left( \nu \right) }\right)
^{T}+\left( x^{\left( \nu \right) }\right) M_{\widetilde{E}}\left( x^{\left(
E\right) }\right) ^{\ast }\left( M_{E}\right) ^{T}\left( y^{\left( E\right)
}\right) ^{T}\right] ,\hspace{0.2cm}\mbox{for the 2 loop model}.%
\end{array}
\right.
\end{equation}
where, for the two loop model, we have set $m_{S}=\max \left( m_{\eta ^{\pm
}},m_{H_{k}^{\pm }},m_{\sigma }\right) $ and we have assumed $m_{\eta ^{\pm
}}\simeq m_{H_{1}^{\pm }}\simeq m_{H_{2}^{\pm }}$ and $m_{E_{i}}$, $m_{%
\widetilde{E}_{i}}<<m_{\eta ^{\pm }}$, $m_{H_{k}^{\pm }}<<$ $m_{\sigma }$ ($%
i=1,2,3$). Besides that, $\lambda$ stands for the quartic scalar coupling
associated with the interaction $\varepsilon _{ab}\left( \eta ^{\ast
}\right) ^{a}\left( \phi ^{\ast }\right) ^{b}S_{2}^{+}\sigma ^{\ast }+h.c$.
Furthermore, under the aforementioned assumptions, the loop function for the
two loop model takes the form \cite{Herrero-Garcia:2014hfa,McDonald:2003zj}: 
\begin{equation}
J\left( \varkappa \right) =\left\{ 
\begin{array}{l}
1+\frac{3}{\pi ^{2}}\left( \ln ^{2}\varkappa -1\right),\hspace{0.2cm}\text{%
for}\hspace{0.2cm}\varkappa >>1 \\ 
\\ 
1,\hspace{0.2cm}\text{for}\hspace{0.2cm}\varkappa \rightarrow 0.%
\end{array}%
\right. ,  \notag
\end{equation}
and the electrically charged scalars in the interaction and physical basis
are related as: 
\begin{eqnarray}
\left( 
\begin{array}{c}
\eta ^{\pm } \\ 
S_{1}^{\pm } \\ 
S_{2}^{\pm }%
\end{array}%
\right) &=&R_{C}\left( 
\begin{array}{c}
H_{1}^{\pm } \\ 
H_{2}^{\pm } \\ 
H_{3}^{\pm }%
\end{array}%
\right) ,\hspace{1cm}\mbox{for the one loop model},  \notag \\
\left( 
\begin{array}{c}
S_{1}^{\pm } \\ 
S_{2}^{\pm }%
\end{array}%
\right) &=&R\left( 
\begin{array}{c}
H_{1}^{\pm } \\ 
H_{2}^{\pm }%
\end{array}%
\right) =\left( 
\begin{array}{cc}
\cos \theta & -\sin \theta \\ 
\sin \theta & \cos \theta%
\end{array}%
\right) \left( 
\begin{array}{c}
H_{1}^{\pm } \\ 
H_{2}^{\pm }%
\end{array}%
\right) ,\hspace{1cm}\eta ^{\pm }=H^{\pm },\hspace{1cm}%
\mbox{for the two
loop model},
\end{eqnarray}
where $R_{C}$ is a real orthogonal $3\times 3$ matrix. 

From the linear seesaw mass matrix (\ref{Mnufull}), (\ref{eq:Mass-Matr2}) we
have the physical neutrino mass matrices 
%
%

\begin{eqnarray}
M_{\nu }^{\left( 1\right) } &\cong &-\left[ \varepsilon M^{-1}m^{T}+m\left(
M^{T}\right) ^{-1}\varepsilon ^{T}\right] , \\
\label{eq:M-2}
M_{\nu }^{\left( 2\right) } &\cong &-\frac{1}{2}\left( M+M^{T}\right) \left[
1_{2\times 2}-\left( M+M^{T}\right) ^{-2}m_{1}^{T}m_{1}\right] -\left( M+M^{T}\right)
^{-1}m_{1}^{T}m_{1},\hspace{0.7cm}\hspace{0.7cm}m_{1}=m-\varepsilon , \\
\label{eq:M-3}
M_{\nu }^{\left( 3\right) } &\cong &\frac{1}{2}\left( M+M^{T}\right) \left[
1_{2\times 2}-\left( M+M^{T}\right)^{-2}m_{2}^{T}m_{2}\right] +\left( M+M^{T}\right)
^{-1}m_{2}^{T}m_{2},\hspace{0.7cm}\hspace{0.7cm}m_{2}=m+\varepsilon ,
\end{eqnarray}

where $M_{\nu }^{\left( 1\right) }$ corresponds to the active neutrino mass
matrix whereas $M_{\nu }^{\left( 2\right) }$ and $M_{\nu }^{\left( 3\right)
} $ are the sterile neutrino mass matrices. The physical neutrino spectrum
is composed of 3 light active neutrinos and 2 pairs of nearly degenerate
sterile exotic pseudo-Dirac neutrinos $N^{\pm}_{1}$ and $N^{\pm}_{2}$. 
For more details about diagonalization see Appendix \ref{fullnumatrix}.

Assuming that the scalar and fermionic seesaw mediators in the diagrams
Figs. \ref{Neutrino1loopdiagram} and \ref{Neutrino2loopdiagram} have masses
at the scales $m_S$ and $m_E$, respectively, the light active neutrino mass
scale can be estimated for the one and two loop models as follows: 
\begin{equation}  \label{eq:1-2-loops}
m_{\nu }\sim \left\{ 
\begin{array}{l}
\frac{\lambda y^{3}v^{2}fm_E}{16\pi ^{2}m_S^2M}\hspace{0.5cm}%
\mbox{for the
one loop model} \\ 
\\ 
\frac{\lambda^{2}y^{4}v^{2}m_E^{2}}{256\pi^{4}m_S^{2}M}\hspace{0.5cm}%
\mbox{for the two loop
model}.%
\end{array}%
\right.
\end{equation}%
where $y$ is a common coupling of the neutrino Yukawa interactions, whereas $%
\lambda $ is the couplings of the quartic scalar interaction $%
S_{1}^{+}S_{2}^{-}\chi \xi $, whereas $f$ is the trilinear scalar coupling
for the interaction $\eta ^{\dagger }\phi \rho$ (see Figures \ref%
{Neutrino1loopdiagram} and \ref{Neutrino2loopdiagram}). Assuming that $%
y\sim\lambda\sim\mathcal{O}(0.1)$, $f\sim m_E\sim\mathcal{O}(1)$ TeV, $%
m_S\sim\mathcal{O}(10)$ TeV, we find from Eq. (\ref{eq:1-2-loops}) that the
light active neutrino mass scale $m_{\nu}\sim 50$ meV can be reproduced in
the one loop linear seesaw model provided that the mass scale of the heavy
pseudo-Dirac neutrinos satisfies $M\sim\mathcal{O}(10^3)$ TeV. Regarding the
two loop linear seesaw model, the choice $\lambda\sim\mathcal{O}(1)$, $y\sim%
\mathcal{O}(0.1)$, $f\sim m_E\sim\mathcal{O}(1)$ TeV, $m_S\sim\mathcal{O}%
(10) $ TeV allows us to reproduce the light active neutrino mass scale $%
m_{\nu}\sim 50$ meV, for pseudo-Dirac neutrinos with masses around $M\sim 
\mathcal{O}(1)$ TeV.

\section{$g-2$ muon anomaly}
\label{sec:G-2} 
The current experimental data on the anomalous dipole
magnetic moment of the muon $a_{\mu }=(g_{\mu }-2)/2$ show significant
deviation from its SM value 
\begin{eqnarray}
\Delta a_{\mu } &=&a_{\mu }^{\mathrm{exp}}-a_{\mu }^{\mathrm{SM}}=\left(
2.49\pm 0.48\right) \times 10^{-9}\hspace{17mm}\mbox{%
\cite{Hagiwara:2011af,Davier:2017zfy,Blum:2018mom,Keshavarzi:2018mgv,Aoyama:2020ynm,Muong-2:2023cdq}}
\label{eq:a-mu}
\end{eqnarray}

\vspace{-4cm}
\begin{figure}[h]
\resizebox{18cm}{24cm}{\hspace{-1cm}%
\includegraphics{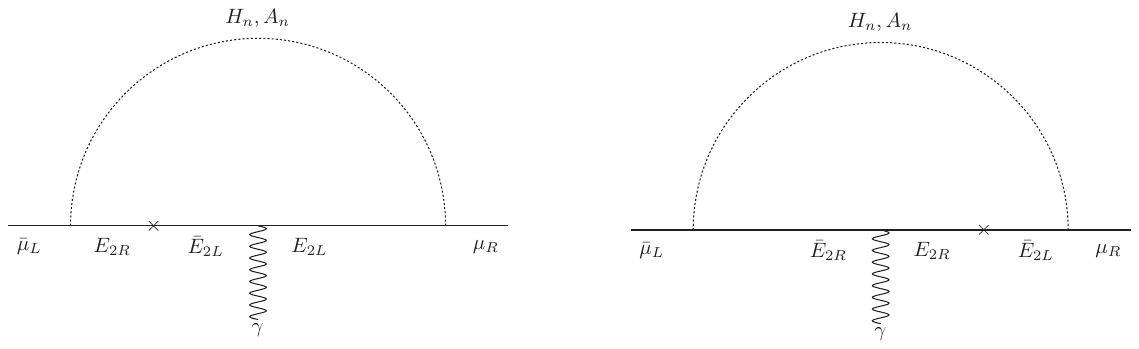}}%
\vspace{-14cm}
\caption{Loop Feynman diagrams contributing to the muon anomalous magnetic
moment. Here $n=1,2$.}
\label{Loopdiagramsgminus2}
\end{figure}

In our models 
this deviation is given by the sum of the partial contributions: 
\begin{equation}
\Delta a_{\mu}=\dsum\limits_{i=1}^{2}\dsum\limits_{\Phi
=H_{i},A_i}\Delta a_{\mu}(\Phi ),
\end{equation}
shown in Fig. \ref{Loopdiagramsgminus2}, which involves the one loop level
exchange of neutral scalars and pseudoscalars as well as charged vector-like
leptons running in the internal lines of the loop. This is different than in
other neutrino mass models, like the ones considered in \cite%
{Hernandez:2021xet,Hue:2021xzl,Cherchiglia:2023utd}, where the muon
anomalous magnetic moment receives contributions arising from the virtual
exchange of electrically charged scalars and right handed Majorana
neutrinos. The analytical form for the neutral scalar and pseudoscalar
contributions at one loop to $\Delta a_{\mu}$ can be found in \cite%
{Diaz:2002uk,Jegerlehner:2009ry,Kelso:2014qka,Lindner:2016bgg,Kowalska:2017iqv}%
. Using these results we write the contributions of the neutral scalars $%
\Phi =H_i,A_i$ ($i=1,2$) defined in Eq. (\ref{Darkscalars}). These
contributions are given by: 
\begin{equation}
\Delta a_{\mu }=\frac{m_{\mu }^{2}}{8\pi ^{2}}\left\{ \dsum\limits_{i=1}^{2}
\omega_{S\mu }^{2}\frac{G_{S}^{\left( \mu \right) }\left(
m_{E_{2}},m_{H_{i}}\right) }{m_{H_{i}}^{2}}+\dsum\limits_{i=1}^{2}
\omega_{P\mu }^{2}\frac{G_{P}^{\left( \mu \right) }\left(
m_{E_{2}},m_{A_{i}}\right) }{m_{A_{i}}^{2}}\right\}  \label{Deltaamu}
\end{equation}


where the loop function is given by: 
\begin{equation}
G_{S,P}^{(l)}(m_{E}, m_{\Phi} )=\int_{0}^{1}dx\frac{x^{2}(1-x\pm \epsilon
_{lE})}{(1-x)(1-x\lambda _{\mu \Phi }^{2})+x\epsilon _{\mu E}^{2}\lambda
_{\mu \Phi}^{2}},\hspace{1cm}\Phi =H_i,A_i  \label{Gloop}
\end{equation}
with $\lambda _{\mu \Phi }=m_{\mu}/m_{\Phi }$, $\epsilon _{\mu
E}=m_{E_{2}}/m_{\mu}$. 
In the loop function $G_{S,P}$, the plus and minus signs stand for
the scalar (CP-even) and pseudoscalar (CP-odd) contributions, respectively.
The quantities $\omega_{S\mu}$ $\omega_{P\mu}$ are the effective Yukawa
couplings for the interactions of the CP-even and CP-odd scalar fields with
fermions in the form $\overline{l}_{iL}\eta E_{jR}$ and $\overline{E}%
_{iL}\rho l_{jR}$ after diagonalization.

The central experimental value of the muon anomalous magnetic moment shown in
Eq. \eqref{eq:a-mu} can be successfully reproduced at the following
benchmark point specified in terms of the masses of the scalars $A_i, H_k$
and charged exotic leptons $E_n$ along with the {effective Yukawa couplings $%
\omega_{S\mu (P\mu)}$: 
\begin{align}  \label{eq:benchmarkmass}
m_{H_1} &\approx 10164.1\,\text{GeV} & m_{H_2} &\approx 3782.1\,\text{GeV} & 
m_{A_1} &\approx 5860.1\,\text{GeV} \\
m_{A_2} &\approx 3781.9\,\text{GeV} & m_{E_1} &\approx 611.5\,\text{GeV} & 
m_{E_2} &\approx 625\,\text{GeV} \\
 \label{eq:benchmarkmass-1} 
\omega_{S\mu} &\approx 1.569 & \omega_{P\mu} &\approx 1.578 & 
\end{align}
According to Eqs.~(\ref{eq:M-eta-rho-1})-(\ref{eq:M-eta-rho-6}), this
benchmark point corresponds to the model parameters 
\begin{align}  \label{eq:benchmarklambda}
v_{\phi} &=246 \,\text{GeV}, & v_{\chi} &\approx 8381.8\,\text{ GeV} \, \  & 
v_{\xi} &\approx 8381.8\ ,\text{GeV} \\
\mu_{\rho} &\approx 383.5\,\text{GeV}\, & \mu_{\eta} &\approx 383.5\, \text{%
GeV} \, & f &\approx 830\,\text{GeV} \\
\lambda_6 &= \lambda_{15}=\lambda_{16} \approx 0.0970 & \lambda_{17}&=%
\lambda_{19}\approx 0.205 & \lambda_{18}&=\lambda_{20}=\lambda_{21}%
\approx0.981
\end{align}
}



\section{Scalar Dark Matter}
\label{section-DM}
As we already mentioned, a viable DM candidate in the one loop radiative linear seesaw model is the 
lightest among the $\widetilde{Z}_{2}$-odd scalars. Regarding the two loop linear seesaw model, due to the preserved $Z_2^{(2)}\times\widetilde{Z}_{2}$ symmetry, one scalar DM candidate is the lightest among the Re $\sigma$ and Im $\sigma$ and the other one is the lightest among $\widetilde{Z}_{2}$-odd scalars. This implies that in this model we have a multicomponent dark matter and thus the resulting relic density will be the sum of the relic densities generated by these two scalar DM candidates.  
As to a fermionic DM candidate, this could be the lightest of the $\widetilde{Z}_{2}$-odd mass eigenstate linear combinations of  $\nu _{nR}$, $N_{nR}$ ($n=1,2$) with the typical mass $M$ specified in Eqs. (\ref{eq:M-2}), (\ref{eq:M-3}).  
However, it is less interesting DM candidate. In fact, as we showed in sec.~\ref{sec:NuMass}, the smallness of the active neutrino masses in Eq. (\ref{eq:1-2-loops}) implies $M$ to be sufficiently large. Otherwise Yukawa and quartic  couplings are very small making the related phenomenology scarce, such as for example very tiny rates for charged lepton flavor violating decays. On the other hand the $\widetilde{Z}_{2}$-odd right handed neutrino states with  $M$ greater than the SM Higgs boson mass $M_H$ are not DM candidates since they can decay into the SM Higgs boson and an active neutrino. 

\begin{figure}[H]
\begin{minipage}{.3333\textwidth}
  \centering
  \includegraphics[width=.8\linewidth]{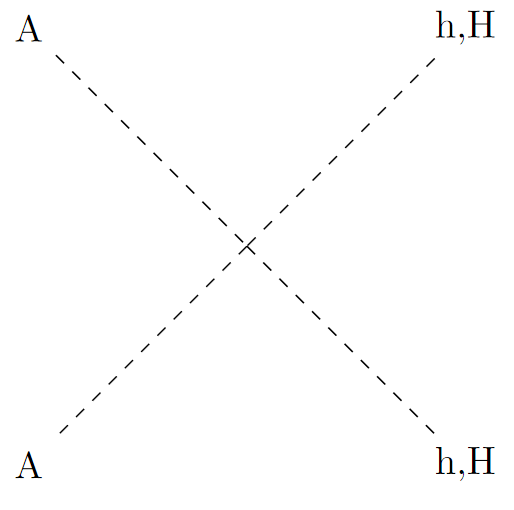}
    \begin{center}
        (a)
    \end{center}
\end{minipage}%
\begin{minipage}{.3333\textwidth}
  \centering
  \includegraphics[width=.8\linewidth]{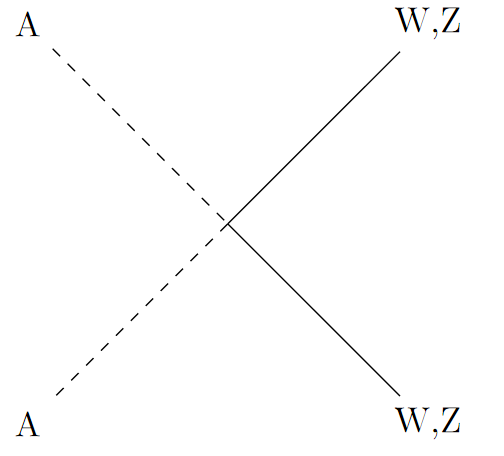}
  \begin{center}
        (b)
    \end{center}
\end{minipage}
\begin{minipage}{.3333\textwidth}
    \centering
    \includegraphics[width=1\linewidth]{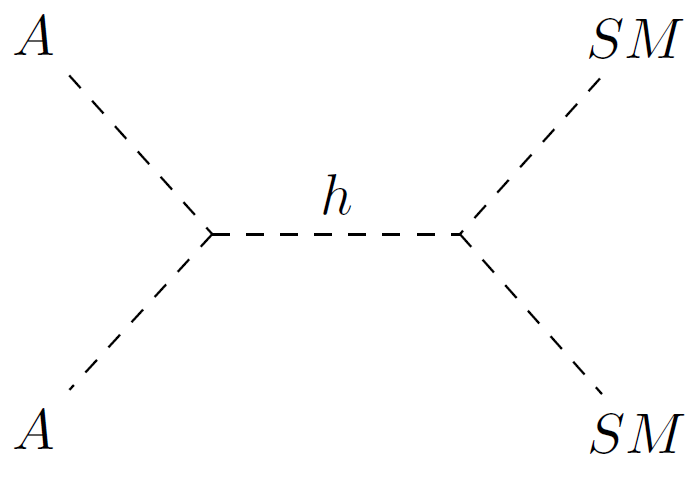}
    \begin{center}
        (c)
    \end{center}
\end{minipage}
\caption{Relevant Feynman diagrams for DM annihilation (a), (b), (c). Here  $A=A_2$ is the CP-odd scalar DM candidate and the lightest CP-even scalar $H=H_2$.}
\label{fig:feynmanDM}
\end{figure}

For these reasons we focus on the scalar DM candidates and proceed to analyze the implications of the one loop linear seesaw model in dark matter. The detailed study of the implications of the two loop linear seesaw model in dark matter is left beyond the scope of the present paper and is deferred for a future work.
%
%
In the benchmark scenario (\ref{eq:benchmarkmass})-(\ref{eq:benchmarklambda}) the DM candidate is the  CP-odd scalar $A_2$, defined in Eq. (\ref{Darkscalars}). We denote this particle as $A$.
After diagonalizing the CP-odd scalar mass matrix in Eq.~(\ref{Metarho}), we find the couplings of 
$A$ to the $126$ GeV SM-like Higgs boson
\begin{align}
	\lambda_{hhAA} =& \lambda_{16} \sin^2\theta_{A}+\lambda_6 \cos ^2\theta_{A}+\lambda _{15} \cos^2\theta_{A} \label{eq:lambdahhAA} \\
   \lambda_{hAA} =& -\sqrt{2} f \sin\theta_{A} \cos\theta_{A} +\lambda_{16} v_{\phi }\sin^2\theta_{A} +\lambda_6 v_{\phi}\cos^2\theta_{A} \nonumber \\
   &+\lambda_{15}v_{\phi } \cos^2\theta_{A} ,
   \label{eq:lambdahAA}
\end{align}
where, $\theta_A\approx 3.13$ is the mixing angle of the CP-odd scalar fields that was fitted with muon g-2 anomaly in Section~\ref{sec:G-2}. 
For the calculation of the relic density abundance $\Omega h^2$ of our scalar DM candidate $A$ we specify its annihilation channels in Fig.~\ref{fig:feynmanDM}. They are 
 annihilation (a) into a pair of the SM Higgses $h$ via quartic coupling  $\lambda_{hhAA}$,  (b) into a pair of the SM electroweak bosons via the gauge coupling as well as (c) 
%
into a pair the SM fermions or the SM Higgses via  a Higgs-portal like diagrams consisting of the trilinear coupling $\lambda_{hAA}$ and the SM Yukawa coupling or trilinear Higgs coupling.
Note that due to the conservation of CP in the scalar sector of our models, there is no coannihilation of $A$ with other scalars, in particular, with $H_2$, which in the current benchmark scenario
(\ref{eq:benchmarkmass})-(\ref{eq:benchmarkmass-1}), is very close in mass to the mass of $A$.


\begin{figure}[H]
     \includegraphics[width=16.0cm, height=12cm]{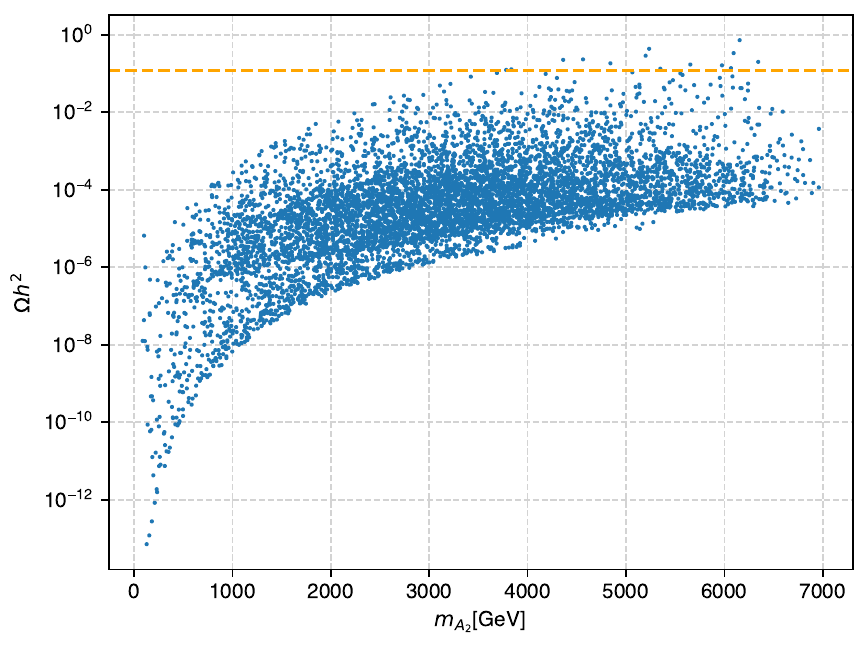}
    \caption{Correlation plot of the relic abundance $\Omega h^2$ as a function of the DM candidate mass $m_{A_2}$. The orange doted line is the experimental limit for Dark Matter relic abundance.}
    \label{fig:corellationdm1}
    \end{figure}

Having this at hand, we calculated DM relic abundance $\Omega h^2$  with the help of micrOMEGAs5.2 \cite{B_langer_2018} and generated a scatter plot characterizing the correlation between $\Omega h^2$ and DM candidate mass $m_{A_2}$, shown in Fig.~\ref{fig:corellationdm1} for the case of the benchmark point (\ref{eq:benchmarkmass})-(\ref{eq:benchmarkmass-1}). 
    \begin{figure}[H]
     \includegraphics[width=16.0cm, height=12cm]{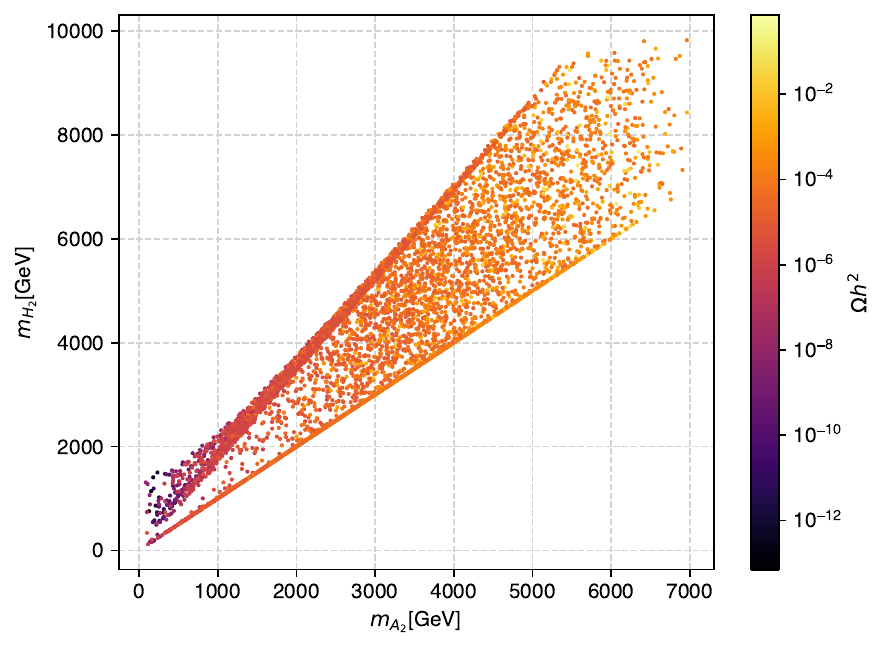}
    \caption{Correlation plot of the relic abundance $\Omega h^2$ as a function of the DM candidate mass $m_{A_2}$ and $m_{H_2}$.}
    \label{fig:corellationdm2}
    \end{figure}


The plot in Fig.~\ref{fig:corellationdm1} shows that the relic abundance is near the observed value $\Omega h^2=0.12\pm 0.0012$ \cite{Planck2020} for $A^0_2$ masses above $4000$ GeV, and only few points are excluded by the limit. Every point in this plot fits $g-2$ muon anomaly. 
The best fit values 
are $\Omega h^2 = 0.1231$ and $\Delta a_{\mu}= 2.79\times 10^{-9}$ 
inside the 
$3\sigma$ and $1\sigma$ experimentally allowed ranges, respectively.

For completeness we studied correlations between the masses of $m_{H_2}$, $m_{A_2}$ and the relic abundance $\Omega h^2$. The result is shown in
Fig.~\ref{fig:corellationdm2}. As seen, the masses of $m_{H_2}$ and $m_{A_2}$ are rather close and linearly correlated. 
Under-abundance is restricted to masses below $2$ TeV. 
The observed relic abundance can be reached in a wide range of larger masses.
 We note that higher values of  $m_{H_2}$ are directly related to the values of relic abundance near the experimental limit and  the values of $m_{A_2}$ are restricted to be 
 close to $m_{H_2}$. 


\section{Charged Lepton Flavor Violation}
\label{clfv}
In this section we analyze charged lepton flavor violation (cLFV) processes
present due to the mixing between active and heavy sterile neutrinos. Here
we focus on $l_i\rightarrow l_j\gamma$ decay. At one loop level its
branching ratio is~\cite{Langacker:1988up,Lavoura:2003xp,Hue:2017lak} 
\begin{eqnarray}
\text{BR}\left( l_{i}\rightarrow l_{j}\gamma \right) &=&\frac{\alpha
_{W}^{3}s_{W}^{2}m_{l_{i}}^{5}}{256\pi ^{2}m_{W}^{4}\Gamma _{i}}\left\vert
G_{ij}\right\vert ^{2}  \label{Brmutoegamma1} \\
G_{ij} &\simeq &\sum_{k=1}^{3}\left( \left[ \left( 1-RR^{\dagger }\right)
U_{\nu }\right] ^{\ast }\right) _{ik}\left( \left( 1-RR^{\dagger }\right)
U_{\nu }\right) _{jk}G_{\gamma }\left( \frac{m_{\nu _{k}}^{2}}{m_{W}^{2}}%
\right) +2\sum_{l=1}^{2}\left( R^{\ast }\right) _{il}\left( R\right)
_{jl}G_{\gamma }\left( \frac{m_{N_{R_{l}}}^{2}}{m_{W}^{2}}\right) ,
\label{Brmutoegamma2} \\
G_{\gamma }(x) &=&\frac{10-43x+78x^{2}-49x^{3}+18x^{3}\ln x+4x^{4}}{12\left(
1-x\right) ^{4}},  \notag
\end{eqnarray}%
where $\Gamma _{\mu }=3\times 10^{-19}$ GeV is the total muon decay width, $%
U_{\nu }$ is the matrix that diagonalizes the light neutrino mass matrix
which, in our case, is equal to the Pontecorvo-Maki-Nakagawa-Sakata (PMNS)
matrix since the charged lepton mixing matrix is set to be equal to the
identity $U_{\ell }=\mathbb{I}$. In addition, the matrix $R$ is given by 
\begin{equation}
R=\frac{1}{\sqrt{2}}mM^{-1},  \label{eq:Rneutrino}
\end{equation}
where $M$ and $m$ are $2\times 2$ and $3\times 2 $ entries of $7\times 7$
neutrino mass matrix in Eq.~(\ref{Mnufull}), respectively. In Fig.~\ref%
{fig:muegamma} we plot the values of the branching ratio Br($\mu \rightarrow
e\gamma$) as a function of $\left|R_{e\mu}\right|$ for different values of Tr%
$[RR^{\dagger}]$. %
%
\begin{figure}[]
\centering
\includegraphics{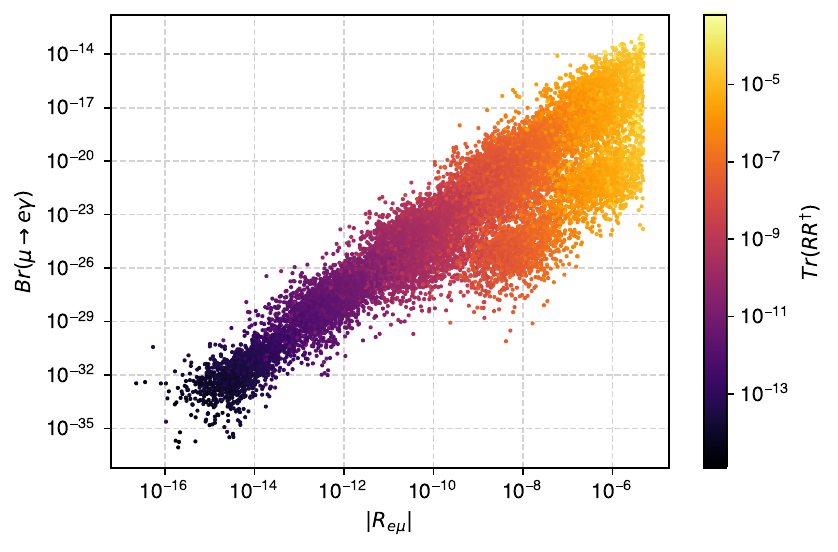}
\caption{$\protect\mu \rightarrow e \protect\gamma$ branching ratio as a
function of $|R_{e\protect\mu}|$. The color surface is the value of Tr$%
[RR^{\dagger}]$.}
\label{fig:muegamma}
\end{figure}
The points have been generated for the benchmark scenario (\ref%
{eq:benchmarkmass})-(\ref{eq:benchmarklambda}) by randomly varying the neutrino
mass matrix parameters (\ref{eq:Mass-Matr2}) in the ranges $m\in [10^{-3},
10^{-1}]$GeV, $M\in [10,10^3]$GeV. According to Sec. \ref{sec:NuMass} these
points are compatible with the active neutrino mass scales $m_{\nu}\sim 50$
meV. Fig.~\ref{fig:muegamma} shows that there are large number of the model
points below the experimental upper bound $Br(\mu \rightarrow
e\gamma)<4.2\times10^{-13}$ many of which should be accessible in the future
experiments.


\section{Leptogenesis}
\label{lepto} 
\begin{figure}[]
\centering
\includegraphics[width=15.0cm, height=12cm]{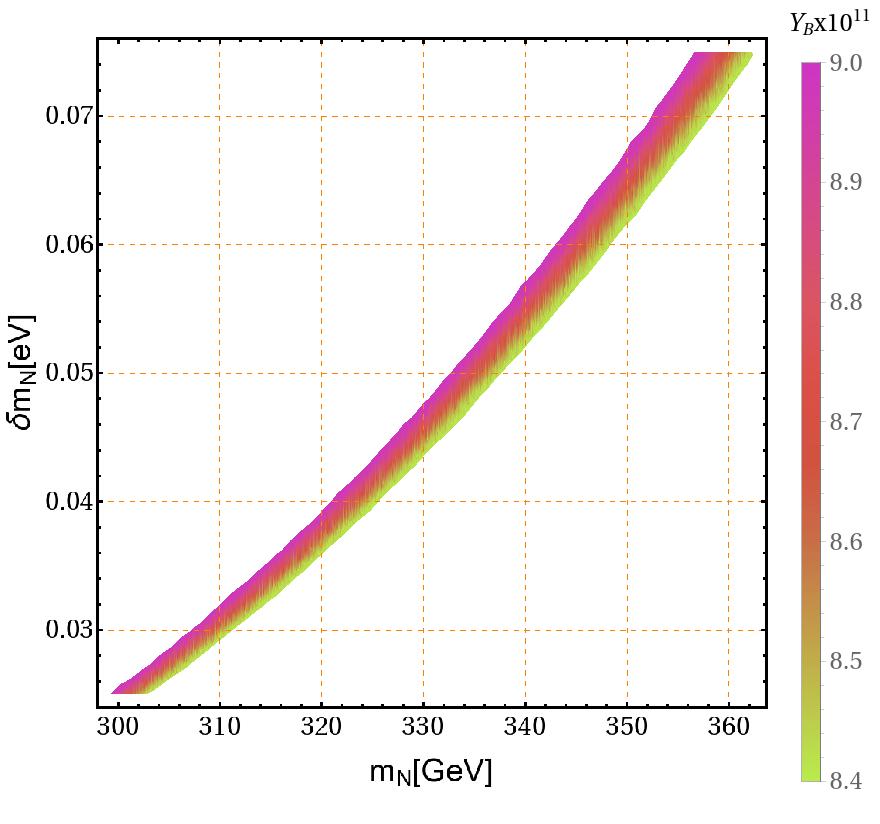}
\caption{Correlation of the sterile neutrino mass splitting $\protect\delta %
m_{N}$ with the mass of the lighest sterile neutrino for the weak washout
regime.}
\label{YB}
\end{figure}

Here we analyze the implications of our two-loop model for leptogenesis. We
skip the one-loop model, where the Yukawa couplings responsible of the
neutrino mass, discussed in Sec.~\ref{sec:NuMass}, are typically smaller
than in the two-loop model and, as a result, its effect on leptogenesis
should be smaller. 
To simplify our analysis we assume that $M$ is a diagonal matrix satisfying
the condition $\left\vert M_{11}\right\vert \ll \left\vert M_{22}\right\vert$%
. In this case 
the Baryon asymmetry of the Universe (BAU) is dominated in our model by $%
N_{1}^{\pm }$ pseudo-Dirac sterile neutrino, defined in Sec. \ref{sec:NuMass}%
. This situation is similar as in \cite{Hernandez:2021uxx,Patel:2023voj}
where the contribution to the baryon asymmetry arising from the remaining
heavy neutral leptons is negligible. 
%
We further assume that the exotic leptonic fields $E_{nR}$ and $\Omega _{nR}$
are heavier than the lightest pseudo-Dirac fermion $N_{1}^{\pm }=N_{\pm }$. 
We use 
the basis where the SM charged lepton mass matrix is diagonal. Then, the
lepton asymmetry parameter, which is induced by the decay of $N_{\pm}$, has
the following form \cite{Gu:2010xc,Pilaftsis:1997jf}:

\begin{equation}
\epsilon_\pm = \frac{\dsum\limits_{i=1}^{3}\left[ \Gamma \left( N_{\pm
}\rightarrow \widetilde{E}_{i}S_{1}^{+}\right) -\Gamma \left( N_{\pm
}\rightarrow \overline{\widetilde{E}}_{i}S_{1}^{-}\right) \right] }{%
\dsum\limits_{i=1}^{3}\left[ \Gamma \left( N_{\pm }\rightarrow \widetilde{E}%
_{i}S_{1}^{+}\right) +\Gamma \left( N_{\pm }\rightarrow \overline{\widetilde{%
E}}_{i}S_{1}^{-}\right) \right] }\simeq \frac{\func{Im}\left\{ \left( \left[
\left( y_{N_{+}}\right) ^{\dagger }\left( y_{N_{-}}\right) \right]
^{2}\right) _{11}\right\} }{8\pi A_{\pm }}\frac{r}{r^{2}+\frac{\Gamma _{\pm
}^{2}}{m_{N_{\pm }}^{2}}},  \label{ep}
\end{equation}%
with: 
\begin{eqnarray}
r &=&\frac{m_{N_{+}}^{2}-m_{N_{-}}^{2}}{m_{N_{+}}m_{N_{-}}},\hspace{0.7cm}%
\hspace{0.7cm}A_{\pm }=\left[ \left( y_{N_{\pm }}\right) ^{\dagger
}y_{N_{\pm }}\right] _{11},\hspace{0.7cm}\hspace{0.7cm}\Gamma _{\pm }=\frac{%
A_{\pm }m_{N_{\pm }}}{8\pi },  \notag \\
y_{N_{+}} &=&\frac{x^{\left( \nu \right) }}{\sqrt{2}}\left( 1_{2\times 2}-%
\frac{1}{2}B_{2}^{T}B_{2}-B_{1}^{T}B_{2}\right) ,  \notag \\
y_{N_{-}} &=&\frac{x^{\left( \nu \right) }}{\sqrt{2}}\left( 1_{2\times 2}-%
\frac{1}{2}B_{1}^{T}B_{1}\right) ,  \label{yN} \\
B_{1} &\simeq &-\sqrt{2}m_{2}\left( M+M^{T}\right) ^{-1}\simeq -\frac{1}{%
\sqrt{2}}m_{2}M^{-1}=-\frac{1}{\sqrt{2}}\left( m-\varepsilon \right) M^{-1},
\notag \\
B_{2} &\simeq &-\sqrt{2}m_{1}\left( M+M^{T}\right) ^{-1}\simeq -\frac{1}{%
\sqrt{2}}m_{1}M^{-1}=-\frac{1}{\sqrt{2}}\left( m+\varepsilon \right) M^{-1}.
\notag
\end{eqnarray}
As mentioned in \cite{Dolan:2018qpy}, neglecting the interference terms
involving the two different sterile neutrinos $N_\pm $ leads to huge values
of the washout parameter $K_{N^{+}}+K_{N^{-}}$. Thus, this effect must be
taken into account. The small mass splitting in the pair of the states $%
N_\pm $, forming a pseudo-Dirac neutrino, leads to destructive interference
in the scattering process \cite{Blanchet:2009kk} properly reducing the
washout parameter. Its effective value, including the interference term, is
given by 
\begin{equation}
K^{eff}\simeq \left( K_{N^{+}}\delta _{+}^{2}+K_{N^{-}}\delta
_{-}^{2}\right) ,
\end{equation}%
where: 
\begin{equation}
\delta _{\pm }=\frac{m_{N^{+}}-m_{N^{-}}}{\Gamma _{N^{\pm }}},\hspace{0.7cm}%
\hspace{0.7cm}K_{N^{\pm }}=\frac{\Gamma _{\pm }}{H\left( T\right) },\hspace{%
0.7cm}\hspace{0.7cm}H(T)=\sqrt{\frac{4\pi ^{3}g^{\ast }}{45}}\frac{T^{2}}{%
M_{P}}
\end{equation}%
where $g^{\ast }=106.75$ is the number of effective relativistic degrees of
freedom, $M_{Pl}=1.2\times 10^{9}$ GeV is the Planck constant and $%
T=m_{N_{\pm }}$. In the weak and strong washout regimes, the baryon
asymmetry is related to the lepton asymmetry \cite{Pilaftsis:1997jf} as
follows 
\begin{eqnarray}
Y_{\Delta B} &=&\frac{n_{B}-\overline{n}_{B}}{s}=-\frac{28}{79}\frac{%
\epsilon _{+}+\epsilon _{-}}{g^{\ast }},\hspace*{0.5cm}\text{for}\hspace*{%
0.5cm}K^{eff}\ll 1, \\
Y_{\Delta B} &=&\frac{n_{B}-\overline{n}_{B}}{s}=-\frac{28}{79}\frac{%
0.3\left( \epsilon _{+}+\epsilon _{-}\right) }{g^{\ast }K^{eff}\left( \ln
K^{eff}\right) ^{0.6}},\hspace*{0.5cm}\text{for}\hspace*{0.5cm}K^{eff}\gg 1,
\end{eqnarray}
The correlation of the sterile neutrino mass splitting $\delta m_{N}$ with
the mass of the lightest sterile neutrino for the weak washout regime is
shown in Figure \ref{YB}. To generate this plot, we performed a random scan
in the ranges $\left(m_{\nu D}\right)_{ij}\in[1, 150]$ GeV, $M_{11}\in[150,
500]$ GeV, $M_{22}\in[1, 2]$ TeV and $\delta m_{N}\in[0.5, 1.5]$ eV. As
indicated by Figure \ref{YB}, an increase of the sterile neutrino mass
splitting $\delta m_{N}$ yields larger values for the mass $m_N$ of the
lightest pseudo-Dirac lepton, in order to successfully reproduce 
the observed value of the baryon asymmetry \cite{Planck:2018vyg} 
\begin{equation}  \label{eq:BAU-observ-1}
Y_{\Delta B}=\frac{n_B-n_{\bar{B}}}{s}=\left( 0.87\pm 0.01\right) \times
10^{-10}
\end{equation}
%
Such larger values of the lightest pseudo-Dirac lepton mass $m_N$ will give
rise to smaller active-sterile neutrino mixing angles, thus yielding smaller
rates for charged lepton flavor violating decays such as $\mu\to e\gamma$.
As shown by Figure \ref{YB}, our model is capable of successfully
accommodating the baryon asymmmetry of the Universe via resonant
leptogenesis. 

\section{Conclusions}

\label{conclusions}

We have proposed two models where the tiny masses of the light active
neutrinos are generated from a radiative linear seesaw mechanism, with the
Dirac neutrino mass submatrix arising at one and two loop levels in the
first and second model, respectively, due to the virtual electrically
charged scalars and vector-like leptons running inside the loops. In these
models, the masses of the SM charged leptons are generated from a one loop
radiative seesaw mechanism, mediated by charged exotic vector like leptons
and electrically neutral scalars. 
This loop pattern, engendering small masses to the light active neutrinos
and the SM charged lepton masses, is ensured by a preserved $\widetilde{Z}_2$
discrete symmetry, which also guarantees stability of the scalar dark matter
candidate in our models. 
%
These models can be treated as 
extended Inert Doublet Models (IDM), where the scalar content is augmented
with the inclusion of several electrically neutral and electrically charged
scalar singlets, whereas the lepton sector is enlarged with right handed
Majorana neutrinos and charged exotic vector like leptons. We have found
that these models can successfully comply with the constraints arising from
charged lepton flavor violation, leptogenesis, dark matter and muon
anomalous magnetic moment.

\section*{Comment}

Our friend and collaborator Iv\'an Schmidt passed away during the completion
of this work. He will be sorely missed.

\section*{Acknowledgments}

A.E.C.H, I.S, S.K and N.P are supported by ANID-Chile FONDECYT 1210378,
ANID-Chile FONDECYT 1180232, ANID-Chile FONDECYT 3150472, ANID-Chile
FONDECYT 1230160, ANID PIA/APOYO AFB230003, Milenio-ANID-ICN2019\_044,
ANID-Chile Doctorado Nacional año 2022 21221396 and Programa de
Incentivo a la Investigaci\'on Cient\'{\i}fica
(PIIC) from UTFSM.

\appendix

\section{Diagonalization of the neutrino mass matrix.}

\label{fullnumatrix} Here, for the convenience of the reader, 
we show in full detail the perturbative diagonalization procedure of the
full $7\times 7$ neutrino mass matrix $M_{\nu }$ of Eq.~(\ref{Mnufull}). The
elements of the submatrices $\varepsilon $, $m$ and $M$ obey the following
hierarchy: 
\begin{equation}
\varepsilon _{in}<<m_{in}<<M_{np},\hspace{0.7cm}\hspace{0.7cm}i=1,2,3,%
\hspace{0.7cm}\hspace{0.7cm}n,p=1,2.  \label{Hierarchy}
\end{equation}

We start by applying the following first orthogonal transformation to the
matrix $M_{\nu }$:

\begin{eqnarray}
S_{\nu }^{T}M_{\nu }S_{\nu } &=&\left( 
\begin{array}{ccc}
1_{3\times 3} & 0_{3\times 2} & 0_{3\times 2} \\ 
0_{2\times 3} & \frac{1}{\sqrt{2}}1_{2\times 2} & -\frac{1}{\sqrt{2}}%
1_{2\times 2} \\ 
0_{2\times 3} & \frac{1}{\sqrt{2}}1_{2\times 2} & \frac{1}{\sqrt{2}}%
1_{2\times 2}%
\end{array}%
\right) \left( 
\begin{array}{ccc}
0_{3\times 3} & \varepsilon & m \\ 
\varepsilon ^{T} & 0_{2\times 2} & M \\ 
m^{T} & M^{T} & 0_{2\times 2}%
\end{array}%
\right) \left( 
\begin{array}{ccc}
1_{3\times 3} & 0_{3\times 2} & 0_{3\times 2} \\ 
0_{2\times 3} & \frac{1}{\sqrt{2}}1_{2\times 2} & \frac{1}{\sqrt{2}}%
1_{2\times 2} \\ 
0_{2\times 3} & -\frac{1}{\sqrt{2}}1_{2\times 2} & \frac{1}{\sqrt{2}}%
1_{2\times 2}%
\end{array}%
\right)  \notag \\
&=&\left( 
\begin{array}{ccc}
1_{3\times 3} & 0_{3\times 2} & 0_{3\times 2} \\ 
0_{2\times 3} & \frac{1}{\sqrt{2}}1_{2\times 2} & -\frac{1}{\sqrt{2}}%
1_{2\times 2} \\ 
0_{2\times 3} & \frac{1}{\sqrt{2}}1_{2\times 2} & \frac{1}{\sqrt{2}}%
1_{2\times 2}%
\end{array}%
\right) \left( 
\begin{array}{ccc}
0_{3\times 3} & -\frac{1}{\sqrt{2}}\left( m-\varepsilon \right) & \frac{1}{%
\sqrt{2}}\left( m+\varepsilon \right) \\ 
\varepsilon ^{T} & -\frac{1}{\sqrt{2}}M & \frac{1}{\sqrt{2}}M \\ 
m^{T} & \frac{1}{\sqrt{2}}M^{T} & \frac{1}{\sqrt{2}}M^{T}%
\end{array}%
\right)  \notag \\
&=&\allowbreak \left( 
\begin{array}{ccc}
0_{3\times 3} & -\frac{1}{\sqrt{2}}\left( m-\varepsilon \right) & \frac{1}{%
\sqrt{2}}\left( m+\varepsilon \right) \\ 
\frac{1}{\sqrt{2}}\left( \varepsilon ^{T}-m^{T}\right) & -\frac{1}{2}\left(
M+M^{T}\right) & \frac{1}{2}M-\frac{1}{2}M^{T} \\ 
\frac{1}{\sqrt{2}}\left( \varepsilon ^{T}+m^{T}\right) & \frac{1}{2}M^{T}-%
\frac{1}{2}M & \frac{1}{2}\left( M+M^{T}\right)%
\end{array}%
\right)  \notag \\
&\cong &\left( 
\begin{array}{ccc}
0_{3\times 3} & -\frac{1}{\sqrt{2}}m_{1} & \frac{1}{\sqrt{2}}m_{2} \\ 
-\frac{1}{\sqrt{2}}m_{1}^{T} & -\frac{1}{2}\left( M+M^{T}\right) & 
0_{2\times 2} \\ 
\frac{1}{\sqrt{2}}m_{2}^{T} & 0_{2\times 2} & \frac{1}{2}\left(
M+M^{T}\right)%
\end{array}%
\right) \allowbreak  \label{T1}
\end{eqnarray}

where the rotation matrix $S_{\nu }$ is given by:

\begin{equation}
S_{\nu }=\left( 
\begin{array}{ccc}
1_{3\times 3} & 0_{3\times 2} & 0_{3\times 2} \\ 
0_{2\times 3} & \frac{1}{\sqrt{2}}1_{2\times 2} & \frac{1}{\sqrt{2}}%
1_{2\times 2} \\ 
0_{2\times 3} & -\frac{1}{\sqrt{2}}1_{2\times 2} & \frac{1}{\sqrt{2}}%
1_{2\times 2}%
\end{array}%
\right) ,  \label{Snu}
\end{equation}

and the submatrices $m_{1}$ and $m_{2}$ have the form:

\begin{equation}
m_{1}=m-\varepsilon ,\hspace{0.7cm}\hspace{0.7cm}m_{2}=m+\varepsilon .
\label{m1andm2}
\end{equation}

Now we perform a second orthogonal transformation under the matrix $M_{\nu }$
as follows:

\begin{eqnarray}
&&R_{1\nu }^{T}S_{\nu }^{T}M_{\nu }S_{\nu }R_{1\nu }  \label{T2} \\
&=&\left( 
\begin{array}{ccc}
1_{3\times 3}-\frac{1}{2}B_{1}B_{1}^{T} & 0_{3\times 2} & B_{1} \\ 
0_{2\times 3} & 1_{2\times 2} & 0 \\ 
-B_{1}^{T} & 0 & 1_{2\times 2}-\frac{1}{2}B_{1}^{T}B_{1}%
\end{array}%
\right) \left( 
\begin{array}{ccc}
0_{3\times 3} & -\frac{1}{\sqrt{2}}m_{1} & \frac{1}{\sqrt{2}}m_{2} \\ 
-\frac{1}{\sqrt{2}}m_{1}^{T} & -\frac{1}{2}\left( M+M^{T}\right) & 
0_{2\times 2} \\ 
\frac{1}{\sqrt{2}}m_{2}^{T} & 0_{2\times 2} & \frac{1}{2}\left(
M+M^{T}\right)%
\end{array}%
\right) \allowbreak  \notag \\
&&\times \left( 
\begin{array}{ccc}
1_{3\times 3}-\frac{1}{2}B_{1}B_{1}^{T} & 0_{3\times 2} & -B_{1} \\ 
0_{2\times 3} & 1_{2\times 2} & 0_{2\times 2} \\ 
B_{1}^{T} & 0_{2\times 2} & 1_{2\times 2}-\frac{1}{2}B_{1}^{T}B_{1}%
\end{array}%
\right)  \notag \\
&=&\left( 
\begin{array}{ccc}
1_{3\times 3}-\frac{1}{2}B_{1}B_{1}^{T} & 0_{3\times 2} & B_{1} \\ 
0_{2\times 3} & 1_{2\times 2} & 0_{2\times 2} \\ 
-B_{1}^{T} & 0_{2\times 2} & 1_{2\times 2}-\frac{1}{2}B_{1}^{T}B_{1}%
\end{array}%
\right) \left( 
\begin{array}{ccc}
\frac{1}{\sqrt{2}}m_{2}B_{1}^{T} & -\frac{1}{\sqrt{2}}m_{1} & \frac{1}{\sqrt{%
2}}m_{2} \\ 
-\frac{1}{\sqrt{2}}m_{1}^{T} & -\frac{1}{2}\left( M+M^{T}\right) & \frac{1}{%
\sqrt{2}}m_{1}^{T}B_{1} \\ 
\frac{1}{\sqrt{2}}m_{2}^{T}+\frac{1}{2}\left( M+M^{T}\right) B_{1}^{T} & 
0_{2\times 2} & \frac{1}{2}\left( M+M^{T}\right)%
\end{array}%
\right) \allowbreak  \notag \\
&\simeq &\left( 
\begin{array}{ccc}
\frac{1}{\sqrt{2}}m_{2}B_{1}^{T}+B_{1}\left( \frac{1}{\sqrt{2}}m_{2}^{T}+%
\frac{1}{2}\left( M+M^{T}\right) B_{1}^{T}\right) & -\frac{1}{\sqrt{2}}m_{1}
& \frac{1}{\sqrt{2}}m_{2}+\frac{1}{2}B_{1}\left( M+M^{T}\right) \\ 
-\frac{1}{\sqrt{2}}m_{1}^{T} & -\frac{1}{2}\left( M+M^{T}\right) & \frac{1}{%
\sqrt{2}}m_{1}^{T}B_{1} \\ 
\frac{1}{\sqrt{2}}m_{2}^{T}+\frac{1}{2}\left( M+M^{T}\right) B_{1}^{T} & 
\frac{1}{\sqrt{2}}B_{1}^{T}m_{1} & \frac{1}{2}\left( 1_{2\times 2}-\frac{1}{2%
}B_{1}^{T}B_{1}\right) \left( M+M^{T}\right) -\frac{1}{\sqrt{2}}%
B_{1}^{T}m_{2}%
\end{array}%
\right) \allowbreak  \notag
\end{eqnarray}

By imposing the partial diagonalization condition:

\begin{equation}
\left( R_{1\nu }^{T}S_{\nu }^{T}M_{\nu }S_{\nu }R_{1\nu }\right)
_{in}=\left( R_{1\nu }^{T}S_{\nu }^{T}M_{\nu }S_{\nu }R_{1\nu }\right)
_{ni}=0,\hspace{0.7cm}\hspace{0.7cm}i=1,2,3,\hspace{0.7cm}\hspace{0.7cm}%
n=1,2.  \label{C1}
\end{equation}

we find the following relation:

\begin{equation}
B_{1}\simeq -\sqrt{2}m_{2}\left( M+M^{T}\right) ^{-1}  \label{B1}
\end{equation}

which implies that:

\begin{eqnarray}
&&R_{1\nu }^{T}S_{\nu }^{T}M_{\nu }S_{\nu }R_{1\nu }  \notag \\
&\simeq &\left( 
\begin{array}{ccc}
-m_{2}\left( M+M^{T}\right) ^{-1}m_{2}^{T} & -\frac{1}{\sqrt{2}}m_{1} & 
0_{3\times 2} \\ 
-\frac{1}{\sqrt{2}}m_{1}^{T} & -\frac{M+M^{T}}{2} & 0_{2\times 2} \\ 
0_{2\times 3} & 0_{2\times 2} & \left[ 1_{2\times 2}-\left( M+M^{T}\right)
^{-1}m_{2}^{T}m_{2}\left( M+M^{T}\right) ^{-1}\right] \frac{\left(
M+M^{T}\right) }{2}+\left( M+M^{T}\right) ^{-1}m_{2}^{T}m_{2}%
\end{array}%
\right) \allowbreak  \notag \\
&=&\left( 
\begin{array}{ccc}
-m_{2}\left( M+M^{T}\right) ^{-1}m_{2}^{T} & -\frac{1}{\sqrt{2}}m_{1} & 
0_{3\times 2} \\ 
-\frac{1}{\sqrt{2}}m_{1}^{T} & X & 0_{2\times 2} \\ 
0_{2\times 3} & 0_{2\times 2} & Y%
\end{array}%
\right) \allowbreak  \label{T2a}
\end{eqnarray}

where $X$ and $Y$ are given by:

\begin{eqnarray}
X &=&-\frac{1}{2}\left( M+M^{T}\right) ,  \notag \\
Y &=&\frac{1}{2}\left[ 1_{2\times 2}-\left( M+M^{T}\right)
^{-1}m_{2}^{T}m_{2}\left( M+M^{T}\right) ^{-1}\right] \left( M+M^{T}\right)
+\left( M+M^{T}\right) ^{-1}m_{2}^{T}m_{2},
\end{eqnarray}

Next, we proceed to apply a third orthogonal transformation obtaining the
following relation:

\begin{eqnarray}
&&R_{2\nu }^{T}R_{1\nu }^{T}S_{\nu }^{T}M_{\nu }S_{\nu }R_{1\nu }R_{2\nu } 
\notag \\
&=&\left( 
\begin{array}{ccc}
1_{3\times 3}-\frac{1}{2}B_{2}B_{2}^{T} & B_{2} & 0_{3\times 2} \\ 
-B_{2}^{T} & 1_{2\times 2}-\frac{1}{2}B_{2}^{T}B_{2} & 0_{2\times 2} \\ 
0_{2\times 3} & 0_{2\times 2} & 1_{2\times 2}%
\end{array}%
\right) \left( 
\begin{array}{ccc}
-m_{2}\left( M+M^{T}\right) ^{-1}m_{2}^{T} & -\frac{1}{\sqrt{2}}m_{1} & 
0_{3\times 2} \\ 
-\frac{1}{\sqrt{2}}m_{1}^{T} & X & 0_{2\times 2} \\ 
0_{2\times 3} & 0_{2\times 2} & Y%
\end{array}%
\right) \allowbreak  \notag \\
&&\times \left( 
\begin{array}{ccc}
1-\frac{1}{2}B_{2}B_{2}^{T} & -B_{2} & 0_{3\times 2} \\ 
B_{2}^{T} & 1_{2\times 2}-\frac{1}{2}B_{2}^{T}B_{2} & 0_{2\times 2} \\ 
0_{2\times 3} & 0_{2\times 2} & 1_{2\times 2}%
\end{array}%
\right)  \notag \\
&\simeq &\left( 
\begin{array}{ccc}
1_{3\times 3}-\frac{1}{2}B_{2}B_{2}^{T} & B_{2} & 0_{3\times 2} \\ 
-B_{2}^{T} & 1_{2\times 2}-\frac{1}{2}B_{2}^{T}B_{2} & 0_{2\times 2} \\ 
0_{2\times 3} & 0_{2\times 2} & 1_{2\times 2}%
\end{array}%
\right) \left( 
\begin{array}{ccc}
-m_{2}\left( M+M^{T}\right) ^{-1}m_{2}^{T}-\frac{1}{\sqrt{2}}m_{1}B_{2}^{T}
& -\frac{1}{\sqrt{2}}m_{1} & 0_{3\times 2} \\ 
-\frac{1}{\sqrt{2}}m_{1}^{T}+XB_{2}^{T} & X & 0_{2\times 2} \\ 
0_{2\times 3} & 0_{2\times 2} & Y%
\end{array}%
\right) \allowbreak  \notag \\
&\simeq &\left( 
\begin{array}{ccc}
-m_{2}\left( M+M^{T}\right) ^{-1}m_{2}^{T}-\frac{1}{\sqrt{2}}m_{1}B_{2}^{T}-%
\frac{1}{\sqrt{2}}B_{2}m_{1}^{T}+B_{2}XB_{2}^{T} & -\frac{1}{\sqrt{2}}%
m_{1}+B_{2}X & 0_{3\times 2} \\ 
-\frac{1}{\sqrt{2}}m_{1}^{T}+XB_{2}^{T} & X\left( 1_{2\times 2}-\frac{1}{2}%
B_{2}^{T}B_{2}\right) +\frac{1}{\sqrt{2}}B_{2}^{T}m_{1} & 0_{2\times 2} \\ 
0_{2\times 3} & 0_{2\times 2} & Y%
\end{array}%
\right) \allowbreak \allowbreak  \label{T3}
\end{eqnarray}

The partial diagonalization condition yields the relation:

\begin{equation}
\left( R_{2\nu }^{T}R_{1\nu }^{T}S_{\nu }^{T}M_{\nu }S_{\nu }R_{1\nu
}R_{2\nu }\right) _{in}=\left( R_{2\nu }^{T}R_{1\nu }^{T}S_{\nu }^{T}M_{\nu
}S_{\nu }R_{1\nu }R_{2\nu }\right) _{ni}=0,\hspace{0.7cm}\hspace{0.7cm}%
i=1,2,3,\hspace{0.7cm}\hspace{0.7cm}n=1,2.  \label{C2}
\end{equation}

then implying the relation:

\begin{equation}
B_{2}\simeq \frac{1}{\sqrt{2}}m_{1}X^{-1}\simeq -\sqrt{2}m_{1}\left(
M+M^{T}\right) ^{-1}.  \label{B2}
\end{equation}

Thus, the $7\times 7$ neutrino mass matrix $M_{\nu }$ of Eq. (\ref{Mnufull})
can be block diagonalized as follows: 
\begin{equation}
R_{2\nu }^{T}R_{1\nu }^{T}S_{\nu }^{T}M_{\nu }S_{\nu }R_{1\nu }R_{2\nu
}\simeq \left( 
\begin{array}{ccc}
M_{\nu }^{\left( 1\right) } & 0_{3\times 2} & 0_{3\times 2} \\ 
0_{2\times 3} & M_{\nu }^{\left( 2\right) } & 0_{2\times 2} \\ 
0_{2\times 3} & 0_{2\times 2} & M_{\nu }^{\left( 3\right) }%
\end{array}%
\right) ,\allowbreak  \label{blockdiagonalization}
\end{equation}

where $M_{\nu }^{\left( 1\right) }$ is the mass matrix for light active
neutrinos, whereas $M_{\nu }^{\left( 2\right) }$ and $M_{\nu }^{\left(
3\right) }$ are the sterile neutrino mass matrices. These matrices are given
by:

\begin{eqnarray}
M_{\nu }^{\left( 1\right) } &\simeq &-m_{2}\left( M+M^{T}\right)
^{-1}m_{2}^{T}-\frac{1}{\sqrt{2}}m_{1}B_{2}^{T}-\frac{1}{\sqrt{2}}%
B_{2}m_{1}^{T}+B_{2}XB_{2}^{T}  \notag \\
&\simeq &-m_{2}\left( M+M^{T}\right) ^{-1}m_{2}^{T}+m_{1}\left(
M+M^{T}\right) ^{-1}m_{1}^{T}+m_{1}\left( M+M^{T}\right)
^{-1}m_{1}^{T}-m_{1}\left( M+M^{T}\right) ^{-1}m_{1}^{T}  \notag \\
&\simeq &-\left[ \varepsilon M^{-1}m^{T}+m\left( M^{T}\right)
^{-1}\varepsilon ^{T}\right] ,
\end{eqnarray}

\begin{eqnarray}
M_{\nu }^{\left( 2\right) } &\simeq &-\frac{1}{2}\left( 1_{2\times 2}-\frac{1%
}{2}B_{2}^{T}B_{2}\right) \left( M+M^{T}\right) +\frac{1}{\sqrt{2}}%
B_{2}^{T}m_{1}  \notag \\
&\simeq &-\frac{1}{2}\left[ 1_{2\times 2}-\left( M+M^{T}\right)
^{-1}m_{1}^{T}m_{1}\left( M+M^{T}\right) ^{-1}\right] \left( M+M^{T}\right)
-\left( M+M^{T}\right) ^{-1}m_{1}^{T}m_{1}
\end{eqnarray}

\begin{eqnarray}
M_{\nu }^{\left( 3\right) } &\simeq &\frac{1}{2}\left( 1_{2\times 2}-\frac{1%
}{2}B_{1}^{T}B_{1}\right) \left( M+M^{T}\right) -\frac{1}{\sqrt{2}}%
B_{1}^{T}m_{2}  \notag \\
&\simeq &\frac{1}{2}\left[ 1_{2\times 2}-\left( M+M^{T}\right)
^{-1}m_{2}^{T}m_{2}\left( M+M^{T}\right) ^{-1}\right] \left( M+M^{T}\right)
+\left( M+M^{T}\right) ^{-1}m_{2}^{T}m_{2}
\end{eqnarray}

\bigskip where:%
\begin{eqnarray}
B_{1} &\simeq &-\sqrt{2}m_{2}\left( M+M^{T}\right) ^{-1}\simeq -\frac{1}{%
\sqrt{2}}m_{2}M^{-1}=-\frac{1}{\sqrt{2}}\left( m-\varepsilon \right) M^{-1},
\\
B_{2} &\simeq &-\sqrt{2}m_{1}\left( M+M^{T}\right) ^{-1}\simeq -\frac{1}{%
\sqrt{2}}m_{1}M^{-1}=-\frac{1}{\sqrt{2}}\left( m+\varepsilon \right) M^{-1}.
\end{eqnarray}

Thus, the rotation matrix $R_{\nu }$ that diagonalizes the full $7\times 7$
neutrino mass matrix $M_{\nu }$ of Eq. (\ref{Mnufull}) has the form:

\begin{eqnarray}
R_{\nu } &=&S_{\nu }R_{1\nu }R_{2\nu }V_{\nu } \\
&\simeq &\left( 
\begin{array}{ccc}
1_{3\times 3} & 0_{3\times 2} & 0_{3\times 2} \\ 
0_{2\times 3} & \frac{1}{\sqrt{2}}1_{2\times 2} & \frac{1}{\sqrt{2}}%
1_{2\times 2} \\ 
0_{2\times 3} & -\frac{1}{\sqrt{2}}1_{2\times 2} & \frac{1}{\sqrt{2}}%
1_{2\times 2}%
\end{array}%
\right) \left( 
\begin{array}{ccc}
1_{3\times 3}-\frac{1}{2}B_{1}B_{1}^{T} & 0_{3\times 2} & -B_{1} \\ 
0_{2\times 3} & 1_{2\times 2} & 0_{2\times 2} \\ 
B_{1}^{T} & 0_{2\times 2} & 1_{2\times 2}-\frac{1}{2}B_{1}^{T}B_{1}%
\end{array}%
\right)  \notag \\
&&\times \left( 
\begin{array}{ccc}
1_{3\times 3}-\frac{1}{2}B_{2}B_{2}^{T} & -B_{2} & 0_{3\times 2} \\ 
B_{2}^{T} & 1_{2\times 2}-\frac{1}{2}B_{2}^{T}B_{2} & 0_{2\times 2} \\ 
0_{2\times 3} & 0_{2\times 2} & 1%
\end{array}%
\right) \left( 
\begin{array}{ccc}
R_{\nu }^{\left( 1\right) } & 0_{3\times 2} & 0_{3\times 2} \\ 
0_{2\times 3} & R_{\nu }^{\left( 2\right) } & 0_{2\times 2} \\ 
0_{2\times 3} & 0_{2\times 2} & R_{\nu }^{\left( 3\right) }%
\end{array}%
\right) ,
\end{eqnarray}

where:

\begin{equation}
V_{\nu }=\left( 
\begin{array}{ccc}
R_{\nu }^{\left( 1\right) } & 0_{3\times 2} & 0_{3\times 2} \\ 
0_{2\times 3} & R_{\nu }^{\left( 2\right) } & 0_{2\times 2} \\ 
0_{2\times 3} & 0_{2\times 2} & R_{\nu }^{\left( 3\right) }%
\end{array}%
\right) ,  \label{Vnu}
\end{equation}

being $R_{\nu }^{\left( 1\right) }$, $R_{\nu }^{\left( 2\right) }$ and $%
R_{\nu }^{\left( 3\right) }$ are the rotation matrices that diagonalize $%
M_{\nu }^{\left( 1\right) }$, $M_{\nu }^{\left( 2\right) }$ and $M_{\nu
}^{\left( 3\right) }$, respectively. The rotation mass matrix $R_{\nu }$
given above, can be rewritten as follows:

\begin{eqnarray}
R_{\nu } &=&S_{\nu }R_{1\nu }R_{2\nu }V_{\nu }  \notag \\
&\simeq &\left( 
\begin{array}{ccc}
1_{3\times 3} & 0_{3\times 2} & 0_{3\times 2} \\ 
0_{2\times 3} & \frac{1}{\sqrt{2}}1_{2\times 2} & \frac{1}{\sqrt{2}}%
1_{2\times 2} \\ 
0_{2\times 3} & -\frac{1}{\sqrt{2}}1_{2\times 2} & \frac{1}{\sqrt{2}}%
1_{2\times 2}%
\end{array}%
\right) \left( 
\begin{array}{ccc}
\left( 1_{3\times 3}-\frac{1}{2}B_{1}B_{1}^{T}\right) \left( 1_{3\times 3}-%
\frac{1}{2}B_{2}B_{2}^{T}\right) & -\left( 1_{3\times 3}-\frac{1}{2}%
B_{1}B_{1}^{T}\right) B_{2} & -B_{1} \\ 
B_{2}^{T} & 1_{2\times 2}-\frac{1}{2}B_{2}^{T}B_{2} & 0_{2\times 2} \\ 
B_{1}^{T}\left( 1_{3\times 3}-\frac{1}{2}B_{2}B_{2}^{T}\right) & 
-B_{1}^{T}B_{2} & 1_{2\times 2}-\frac{1}{2}B_{1}^{T}B_{1}%
\end{array}%
\right)  \notag \\
&&\times \left( 
\begin{array}{ccc}
R_{\nu }^{\left( 1\right) } & 0_{3\times 2} & 0_{3\times 2} \\ 
0_{2\times 3} & R_{\nu }^{\left( 2\right) } & 0_{2\times 2} \\ 
0_{2\times 3} & 0_{2\times 2} & R_{\nu }^{\left( 3\right) }%
\end{array}%
\right)  \notag \\
&\simeq &\left( 
\begin{array}{ccc}
\left( 1_{3\times 3}-\frac{1}{2}B_{1}B_{1}^{T}\right) \left( 1_{3\times 3}-%
\frac{1}{2}B_{2}B_{2}^{T}\right) & -\left( 1_{3\times 3}-\frac{1}{2}%
B_{1}B_{1}^{T}\right) B_{2} & -B_{1} \\ 
\frac{B_{1}^{T}+B_{2}^{T}}{\sqrt{2}} & \frac{1}{\sqrt{2}}\left( 1_{2\times
2}-\frac{1}{2}B_{2}^{T}B_{2}-B_{1}^{T}B_{2}\right) & \frac{1}{\sqrt{2}}%
\left( 1_{2\times 2}-\frac{1}{2}B_{1}^{T}B_{1}\right) \\ 
\frac{B_{1}^{T}-B_{2}^{T}}{\sqrt{2}} & -\frac{1}{\sqrt{2}}\left( 1_{2\times
2}-\frac{1}{2}B_{2}^{T}B_{2}+B_{1}^{T}B_{2}\right) & \frac{1}{\sqrt{2}}%
\left( 1_{2\times 2}-\frac{1}{2}B_{1}^{T}B_{1}\right)%
\end{array}%
\right) \left( 
\begin{array}{ccc}
R_{\nu }^{\left( 1\right) } & 0_{3\times 2} & 0_{3\times 2} \\ 
0_{2\times 3} & R_{\nu }^{\left( 2\right) } & 0_{2\times 2} \\ 
0_{2\times 3} & 0_{2\times 2} & R_{\nu }^{\left( 3\right) }%
\end{array}%
\right)  \notag \\
&\simeq &\left( 
\begin{array}{ccc}
1_{3\times 3} & -B_{2} & -B_{1} \\ 
\frac{B_{1}^{T}+B_{2}^{T}}{\sqrt{2}} & \frac{1}{\sqrt{2}}\left( 1_{2\times
2}-\frac{1}{2}B_{2}^{T}B_{2}-B_{1}^{T}B_{2}\right) & \frac{1}{\sqrt{2}}%
\left( 1_{2\times 2}-\frac{1}{2}B_{1}^{T}B_{1}\right) \\ 
\frac{B_{1}^{T}-B_{2}^{T}}{\sqrt{2}} & -\frac{1}{\sqrt{2}}\left( 1_{2\times
2}-\frac{1}{2}B_{2}^{T}B_{2}+B_{1}^{T}B_{2}\right) & \frac{1}{\sqrt{2}}%
\left( 1_{2\times 2}-\frac{1}{2}B_{1}^{T}B_{1}\right)%
\end{array}%
\right) \left( 
\begin{array}{ccc}
R_{\nu }^{\left( 1\right) } & 0_{3\times 2} & 0_{3\times 2} \\ 
0_{2\times 3} & R_{\nu }^{\left( 2\right) } & 0_{2\times 2} \\ 
0_{2\times 3} & 0_{2\times 2} & R_{\nu }^{\left( 3\right) }%
\end{array}%
\right)  \notag \\
&\simeq &\left( 
\begin{array}{ccc}
R_{\nu }^{\left( 1\right) } & -B_{2}R_{\nu }^{\left( 2\right) } & 
-B_{1}R_{\nu }^{\left( 3\right) } \\ 
\frac{B_{1}^{T}+B_{2}^{T}}{\sqrt{2}}R_{\nu }^{\left( 1\right) } & \frac{1}{%
\sqrt{2}}\left( 1_{2\times 2}-\frac{1}{2}B_{2}^{T}B_{2}-B_{1}^{T}B_{2}%
\right) R_{\nu }^{\left( 2\right) } & \frac{1}{\sqrt{2}}\left( 1_{2\times 2}-%
\frac{1}{2}B_{1}^{T}B_{1}\right) R_{\nu }^{\left( 3\right) } \\ 
\frac{B_{1}^{T}-B_{2}^{T}}{\sqrt{2}}R_{\nu }^{\left( 1\right) } & -\frac{1}{%
\sqrt{2}}\left( 1_{2\times 2}-\frac{1}{2}B_{2}^{T}B_{2}+B_{1}^{T}B_{2}%
\right) R_{\nu }^{\left( 2\right) } & \frac{1}{\sqrt{2}}\left( 1_{2\times 2}-%
\frac{1}{2}B_{1}^{T}B_{1}\right) R_{\nu }^{\left( 3\right) }%
\end{array}%
\right)  \label{Rnu}
\end{eqnarray}

Then, we obtain the relation:

\begin{eqnarray}
R_{\nu } &=&S_{\nu }R_{1\nu }R_{2\nu }V_{\nu }  \notag \\
&\simeq &\left( 
\begin{array}{ccc}
R_{\nu }^{\left( 1\right) } & -B_{2}R_{\nu }^{\left( 2\right) } & 
-B_{1}R_{\nu }^{\left( 3\right) } \\ 
\frac{B_{1}^{T}+B_{2}^{T}}{\sqrt{2}}R_{\nu }^{\left( 1\right) } & \frac{1}{%
\sqrt{2}}\left( 1_{2\times 2}-\frac{1}{2}B_{2}^{T}B_{2}-B_{1}^{T}B_{2}%
\right) R_{\nu }^{\left( 2\right) } & \frac{1}{\sqrt{2}}\left( 1_{2\times 2}-%
\frac{1}{2}B_{1}^{T}B_{1}\right) R_{\nu }^{\left( 3\right) } \\ 
\frac{B_{1}^{T}-B_{2}^{T}}{\sqrt{2}}R_{\nu }^{\left( 1\right) } & -\frac{1}{%
\sqrt{2}}\left( 1_{2\times 2}-\frac{1}{2}B_{2}^{T}B_{2}+B_{1}^{T}B_{2}%
\right) R_{\nu }^{\left( 2\right) } & \frac{1}{\sqrt{2}}\left( 1_{2\times 2}-%
\frac{1}{2}B_{1}^{T}B_{1}\right) R_{\nu }^{\left( 3\right) }%
\end{array}%
\right) .  \label{Rnub}
\end{eqnarray}

On the other hand, using Eq. (\ref{Rnub}) we find that the neutrino fields $%
\nu _{L}=\left( \nu _{1L},\nu _{2L},\nu _{3L}\right) ^{T}$, $\nu
_{R}^{C}=\left( \nu _{1R}^{C},\nu _{2R}^{C}\right) $ and $N_{R}^{C}=\left(
N_{1R}^{C},N_{2R}^{C}\right) $ are related with the physical neutrino fields
by the following relations: 
\begin{eqnarray}
\left( 
\begin{array}{c}
\nu _{L} \\ 
\nu _{R}^{C} \\ 
N_{R}^{C}%
\end{array}%
\right) &=&R_{\nu }\Psi _{L}\simeq \left( 
\begin{array}{ccc}
R_{\nu }^{\left( 1\right) } & -B_{2}R_{\nu }^{\left( 2\right) } & 
-B_{1}R_{\nu }^{\left( 3\right) } \\ 
\frac{B_{1}^{T}+B_{2}^{T}}{\sqrt{2}}R_{\nu }^{\left( 1\right) } & \frac{1}{%
\sqrt{2}}\left( 1_{2\times 2}-\frac{1}{2}B_{2}^{T}B_{2}-B_{1}^{T}B_{2}%
\right) R_{\nu }^{\left( 2\right) } & \frac{1}{\sqrt{2}}\left( 1_{2\times 2}-%
\frac{1}{2}B_{1}^{T}B_{1}\right) R_{\nu }^{\left( 3\right) } \\ 
\frac{B_{1}^{T}-B_{2}^{T}}{\sqrt{2}}R_{\nu }^{\left( 1\right) } & -\frac{1}{%
\sqrt{2}}\left( 1_{2\times 2}-\frac{1}{2}B_{2}^{T}B_{2}+B_{1}^{T}B_{2}%
\right) R_{\nu }^{\left( 2\right) } & \frac{1}{\sqrt{2}}\left( 1_{2\times 2}-%
\frac{1}{2}B_{1}^{T}B_{1}\right) R_{\nu }^{\left( 3\right) }%
\end{array}%
\right) \left( 
\begin{array}{c}
\Psi _{L}^{\left( 1\right) } \\ 
\Psi _{L}^{\left( 2\right) } \\ 
\Psi _{L}^{\left( 3\right) }%
\end{array}%
\right) ,  \notag \\
\Psi _{L} &=&\left( 
\begin{array}{c}
\Psi _{L}^{\left( 1\right) } \\ 
\Psi _{L}^{\left( 2\right) } \\ 
\Psi _{L}^{\left( 3\right) }%
\end{array}%
\right) ,
\end{eqnarray}%
where $\Psi _{jL}^{\left( 1\right) }$, $\Psi _{kL}^{\left( 2\right)
}=N_{k}^{+}$ and $\Psi _{kL}^{\left( 3\right) }=N_{k}^{-}$ ($j=1,2,3$ and $%
k=1,2$) are the three active neutrinos and four exotic neutrinos,
respectively.

\bibliographystyle{utphys}
\bibliography{biblioLS}

\end{document}